%% file: main.tex
\documentclass{article}

\usepackage{microtype}
\usepackage{graphicx}
\usepackage{caption}
\usepackage{subcaption}
\usepackage{booktabs} %

\usepackage[hidelinks]{hyperref}

\usepackage[dvipsnames]{xcolor}
\usepackage{color, colortbl}
\usepackage{mathtools}
\usepackage{makecell}
\usepackage{pgfplots}
\usepackage{tikz} 
\usetikzlibrary{arrows, decorations.markings, shapes, arrows, fit, patterns}
\usepackage{nameref}
\usepackage{listings}
\definecolor{codegreen}{rgb}{0,0.6,0}
\definecolor{codegray}{rgb}{0.5,0.5,0.5}
\definecolor{codepurple}{rgb}{0.58,0,0.82}
\definecolor{backcolour}{rgb}{0.95,0.95,0.92}
\lstdefinestyle{mystyle}{
	backgroundcolor=\color{backcolour},   
	commentstyle=\color{codegreen},
	keywordstyle=\color{magenta},
	numberstyle=\tiny\color{codegray},
	stringstyle=\color{codepurple},
	basicstyle=\ttfamily\footnotesize,
	breakatwhitespace=false,         
	breaklines=true,                 
	captionpos=b,                    
	keepspaces=true,                 
	numbers=left,                    
	numbersep=5pt,                  
	showspaces=false,                
	showstringspaces=false,
	showtabs=false,                  
	tabsize=2
}
\usepackage{pgf-pie}
\pgfplotsset{compat=1.17}

\usepackage{ragged2e}
\usepackage{paralist}
\usepackage{adjustbox}
\usepackage{setspace}
\usepackage[normalem]{ulem}

\definecolor{Layer1}{HTML}{D7191C}
\definecolor{Layer2}{HTML}{FDAE61}
\definecolor{Layer3}{HTML}{ABDDA4}
\definecolor{LSTM}{HTML}{2B83BA}
\definecolor{Layer5}{HTML}{94007E}
\definecolor{Layer6}{HTML}{00E225}

\newcommand\LayerOnePattern{horizontal lines}
\newcommand\LayerTwoPattern{vertical lines}
\newcommand\LayerThreePattern{grid}
\newcommand\LayerLSTMPattern{dots}
\newcommand\LayerFourPattern{north east lines}
\newcommand\LayerFivePattern{horizontal lines}

\definecolor{C1}{HTML}{D7191C}
\definecolor{C2}{HTML}{FDAE61}
\definecolor{C3}{HTML}{ABDDA4}
\definecolor{C4}{HTML}{2B83BA}
\definecolor{C5}{HTML}{94007E}
\definecolor{C6}{HTML}{00E225}
\definecolor{C7}{HTML}{FF0000}
\definecolor{C8}{HTML}{a1a1a1}
\definecolor{C9}{HTML}{ec008b}
\definecolor{C10}{HTML}{00b200}
\definecolor{C11}{HTML}{54c7f4}
\definecolor{C12}{HTML}{7f007f}
\definecolor{C13}{HTML}{ff8710}
\definecolor{C14}{HTML}{329898}

\newcommand\COnePattern{horizontal lines}
\newcommand\CTowPattern{vertical lines}
\newcommand\CThreePattern{grid}
\newcommand\CFourPattern{dots}
\newcommand\CFivePattern{north east lines}
\newcommand\CSixPattern{horizontal lines}
\newcommand\CSevenPattern{horizontal lines}
\newcommand\CEightPattern{grid}
\newcommand\CNinePattern{dots}
\newcommand\CTenPattern{north east lines}
\newcommand\CElevenPattern{horizontal lines}
\newcommand\CTwelvePattern{vertical lines}
\newcommand\CThirteenPattern{grid}
\newcommand\CFourteenPattern{dots}

\usepackage[arxiv]{mlsys2023}

\renewcommand{\algorithmicrequire}{\textbf{Input:}}
\renewcommand{\algorithmicensure}{\textbf{Output:}}

\newcommand{\ourmethod}{\textcolor{red}{\textbf{our method}}}
\newcommand{\ourmethodwa}[1]{\textcolor{red}{\texttt{Ours W#1A#1}}}
\newcommand{\ourmethodWA}[1]{\textcolor{red}{\texttt{Ours #1}}}
\renewcommand{\ourmethod}{\emph{FullPack}}
\renewcommand{\ourmethodwa}[1]{\texttt{FullPack-W#1A#1}}
\renewcommand{\ourmethodWA}[1]{\texttt{FullPack-#1}}
\newcommand{\ruyint}{\texttt{Ruy-W8A8}}
\newcommand{\ruyfp}{\texttt{Ruy-FP32}}
\newcommand{\ruy}{\emph{Ruy}}
\newcommand{\xnnpackint}{\texttt{XNNPack-W8A8}}
\newcommand{\xnnpackfp}{\texttt{XNNPack-FP32}}
\newcommand{\xnnpack}{\emph{XNNPACK}}
\newcommand{\tfliteint}{\texttt{TFLite-W8A8}}
\newcommand{\tflitefp}{\texttt{TFLite-FP32}}
\newcommand{\gemmlowpint}{\texttt{GEMMLOWP-W8A8}}
\newcommand{\gemmlowp}{\emph{GEMMLOWP}}
\newcommand{\eigenfp}{\texttt{Eigen-FP32}}
\newcommand{\eigen}{\emph{Eigen}}
\newcommand{\ullpackwa}[1]{\texttt{ULPPACK$^-$-#1}}
\newcommand{\ullpack}{\emph{ULPPACK}}
\newcommand{\ulppackwa}[1]{\texttt{ULPPACK$^-$-#1}}
\newcommand{\ulppackWA}[1]{\texttt{ULPPACK$^-$-W#1A#1}}

\definecolor{maxGreen}{HTML}{009f02}

\mlsystitlerunning{FullPack: Full Vector Utilization for Sub-Byte Quantized Inference on General Purpose CPUs}

\begin{document}
	
	\twocolumn[
	\mlsystitle{FullPack: Full Vector Utilization for Sub-Byte Quantized Inference on General Purpose CPUs}

	\begin{mlsysauthorlist}
		\mlsysauthor{Hossein~Katebi}{Sharif}
		\mlsysauthor{Navidreza~Asadi}{TUM}
		\mlsysauthor{Maziar~Goudarzi}{Sharif}
	\end{mlsysauthorlist}

	\mlsysaffiliation{TUM}{Computer Engineering Department, Technical University of Munich, Munich, Germany (navidreza.asadi@tum.de). He was with Sharif University of Technology, Tehran, Iran, while working on this project.}
	
	\mlsysaffiliation{Sharif}{Department of Computer Engineering, Sharif University of Technology, Tehran, Iran\\}
	
	\mlsyscorrespondingauthor{Maziar~Goudarzi}{goudarzi@sharif.edu}
	
	\mlsyskeywords{Deep Learning, Hardware Acceleration}
	
	\vskip 0.3in
	
	\input{Sections/abstract}

	]

	\printAffiliationsAndNotice{}  %

	\input{Sections/introduction}

	\input{Sections/relatedwork}

	\input{Sections/method}
	
	\input{Sections/evaluation}

	\input{Sections/conclusion.tex}

	\bibliography{main}
	\bibliographystyle{mlsys2023}

	\cleardoublepage
	\appendix
	\input{Sections/appendix.tex}
\end{document}

%% file: Sections/abstract.tex
\newcommand*{\AbstractIV}{V4-Revised By Professor and added experimental setup}
\begin{abstract}
\label{abstract}
	\justifying
	\ifdefined\AbstractIV
	
	Although prior art has demonstrated negligible accuracy drop in sub-byte quantization---where weights and/or activations
	are represented by less than 8 bits---popular SIMD instructions of CPUs do not natively support these datatypes.
	While recent methods, such as ULPPACK, are already using sub-byte quantization on general-purpose CPUs with vector
	units, they leave out several empty bits between the sub-byte values in memory and in vector registers to avoid overflow 
	to the neighbours during the operations. This results in memory footprint and bandwidth-usage inefficiencies and suboptimal performance.
	In this paper, we present memory layouts for storing, and mechanisms for processing sub-byte (4-, 2-, or 1-bit) models that 
	utilize all the bits in the memory as well as in the vector registers for the actual data.
	We provide compute kernels for the proposed layout for the GEMV (GEneral Matrix-Vector multiplication)
	operations between weights and activations of different datatypes (e.g., 8-bit activations and 4-bit weights).
	For evaluation, we extended the TFLite package and added our methods to it, then ran the models on the cycle-accurate gem5 simulator to 
	compare detailed memory and CPU cycles of each method.
	We compare against nine other methods that are actively used in production including GEMLOWP, Ruy, XNNPack, and ULPPACK.
	Furthermore, we explore the effect of different input and output sizes of deep learning layers on the performance of our proposed
	method.
	Experimental results show $0.96{-}2.1{\times}$ speedup for small sizes and $1.2{-}6.7\times$ speedup for mid to large sizes.
	Applying our proposal to a real-world speech recognition model, Mozilla DeepSpeech, we proved that our method
	achieves $1.56{-}2.11\times$ end-to-end speedup compared to the state-of-the-art, depending on the bit-width employed.
	
	\fi
	\ifdefined\AbstractIII
	
	Quantization is an effective approach for the inference-time efficiency of deep learning. 
	Although recent works have demonstrated a negligible accuracy drop in sub-byte quantization (weights and/or activations are represented by ${<8}$ bits), current mobile SIMD instructions do not natively support these datatypes.
	However, existing methods, such as ULPPACK, are already using sub-byte quantization on general-purpose CPUs with vector units but they mostly leave out several bits between the values to avoid overflow of their operations.
	In this paper, we present a memory layout that utilizes all bits in vectors and their corresponding assembly kernels for storing and processing sub-byte (4-, 2-, or 1-bit) models.
	We design and optimize kernels compatible with the proposed layout for the GEMV (GEneral Matrix-Vector multiplication) operations between weights and activations of different datatypes (e.g., 8-bit activations and 4-bit weights).
	We compare our technique to eight methods that are currently being used in production including GEMLOWP, Ruy, and XNNPack.
	We explore the effect of different input and output sizes of deep learning layers on the performance of the proposed method. We get 0.13$\times$ to 1.7$\times$ speedup for small sizes and 1.57$\times$ to 2$\times$ speedup for mid to large sizes.
	We also evaluate our proposal on a real-world speech recognition model, Mozilla DeepSpeech, and show that our method achieves 1.13$\times$ to 1.52$\times$ end-to-end speedup compared to the state-of-the-art, depending on the bit-width.
	
	\fi
	\ifdefined\AbstractII
	
	Quantization is a common approach for deployment of deep learning models on smartphones as it makes excution of models on resource constrained devices either feasible or faster.
	Recent works demonstrate negligible drop in accuracy of sub-byte quantized models: weights and/or activations are represented by ${<8}$ bits.
	Although one may expect performance improvements when using fewer bits, it is difficult to achieve in practice, mostly because most of current mobile SIMD instructions do not support sub-byte operations.
	In this paper, we present better a memory layout and assembly kernels for storing and processing sub-byte (4-, 2-, or 1-bit) models. %
	We optimized the kernels so to utilize the proposed layout and the GEMV (GEneral Matrix-Vector multiplication) operations.
	Our kernels can process the operations between weights and activations of different data types (e.g., 8-bit activations and 4-bit weights).
	We explore the effect of different input and output sizes of deep learning layers on the performance of the proposed method.
	We found that the speed-up of the proposed method compared to the current state-of-the-art varies between 4\% to 70\% depending on the size of L2.
	We also evaluated our proposal on a real-world speech recognition model, Mozilla DeepSpeech.
	We show that our method can achieve an end-to-end speedup between 13\% to 50\% depending on the bit-width.

	\fi
	\ifdefined\AbstractI
	
	\{V1\}
	
	Quantized Deep Neural Networks (QDNN) are common practice for implementing DNN applications on mobile devices.
	On the other hand, recent works demonstrated a negligible accuracy drop in sub-byte models.
	However, there is a lack of methods for processing sub-byte models on mobile phones optimally.
	Hence, in this paper, we present a novel memory layout and corresponding assembly kernels to store and process sub-byte (4-, 2-, or 1-bit) models efficiently.
	The assembly kernels are optimized to utilize the proposed layout and do the GEMV (GEneral Matrix-Vector multiplication) operations.
	These kernels can also process the GEMV between weights and activations of different data types (for example, 8-bit activations and 4-bit weights).
	First, we explored the effect of changing the input/output size of the DNN layers on the performance of the proposed method.
	In this experimentation, we found that the speed-up of the proposed method against the current state-of-the-art varies between 4\% to 70\% depending on the size of L2 and L3 caches.
	Then we moved on to a real-world Speech Recognition application, DeepSpeech, and investigated the effect of applying the proposed method to the end-to-end performance of the model.
	The proposed method could achieve a speed-up varying between 13\% to 50\% based on the bit-width of the computation.
	
	\fi
	
\end{abstract}

%% file: Sections/introduction.tex
\section{Introduction}
\label{introduction}

	Deep Neural Networks (DNN) are showing promising results in various areas such as Natural Language Processing
	\cite{devlin2018bert}, Speech Recognition \cite{amodei2016deepspeech2,hannun2014deepspeech}, and Computer Vision
	\cite{he2016deepresnet,ma2018shufflenet,sandler2018mobilenetv2}.
	While DNNs are getting more attention every day, they are compute-intensive.
	In the past decade, we have witnessed a significant increase in computation power of modern computers, enabling DNN applications.
	However, running DNN models on mobile devices is still challenging.
	The comunity has introduced different quantization schemes to decrease the required computation demands of DNNs.
	These schemes usually quantize weights and/or activations to 8-bit or 16-bit integers.
	Recent works \cite{gong2019differentiable,jung2019learning,jacob2018googlequantization} demonstrated negligible accuracy loss and 
	considerable performance gains.
	For instance, \cite{esser2019learned} show that reducing precision of weights and activations to sub-byte has little to no impact 
	on accuracy. In fact, it might even have a positive impact \cite{banbury2021micronets} when applying hardware-aware neural architecture 
	search; because in a memory-constrained device, reducing precision of parameters makes room for using more parameters, hence a potential 
	for higher accuracy.
	
	Despite performance gain by sub-byte quantized models in specialized hardware, such models remain infeasible or inefficient on general 
	purpose CPUs due to the lack of support in current popular CPU architectures.
	Recent works \cite{won2022ulppack,reggiani2022bison} present novelties to use sub-byte quantized models on CPUs by either 
	adding special hardwares to the current processor \cite{reggiani2022bison} or leaving some bits unused when storing and processing the data
	so as to avoid overflow among adjacent data elements when executing vector instructions \cite{won2022ulppack}. 
	The former needs hardware change and the latter wastes parts of the available memory capacity and bandwidth, and vector processing capability. 
	Existing production-ready frameworks for mobile devices, such as TensorFlowLite and PyTorch are already using low-precision 
	linear algebra libraries to gain performance for quantized models.
	TensorFlowLite (TFLite) allows to use XNNPack \cite{website:xnnpack}, Ruy \cite{website:ruy}, or GEMMLOWP \cite{website:gemmlowp}
	for quantized models while PyTorch supports QNNPACK \cite{website:qnnpack} and FBGEMM \cite{article:fbgemm}.
	Nevertheless, these libraries do not support sub-byte quantized models.

	We provide mechanisms that enable unmodified off-the-shelf processors, with vector ISA, to use sub-byte quantized models without waste
	of valuable memory and processing capacity observed in state of the art.
	We first propose a memory layout to store sub-byte weights and/or activations.
	Our memory layout is tailored to the vector operations inside the CPU and utilizes the whole space, not leaving even a single bit unused.
	This enables the vector instructions to extract multiple blocks by consecutive single shifts.
	We provide various kernels in ARMv8 assembly language to utilize our memory layout and vector operations schedule so as to efficiently 
	implement GEMV on weights and activations using ARM NEON vector instructions.
	Our performance gains come from better storage and communication (including cache space utilization, memory bandwidth utilization, and 
	memory footprint) as well as improved processing (full utilization of the vector registers and vector processing units) despite some 
	additional instructions needed to extract data from the compressed memory layout.
	We have implemented our method on TensorFlowLite and have made it available open-source as a fork of TensorFlow
	\footnote{\href{https://github.com/shkatebi97/tensorflow}{https://github.com/shkatebi97/tensorflow}} 
  
	\input{figures/mozilla-deepspeech-per-layer-breakdown.tex}

	In \ourmethod{}, we decided to focus on GEMV---which is a special case for GEMM---because it did not get enough improvement as 
	the GEMM but is causing significant performance degradation in models utilizing it, such as RNN and LSTM-based models.
	To support this statement, we evaluated Mozilla DeepSpeech \cite{website:mozilladeepspeech} with different quantization
	schemes and provided a per-layer execution time breakdown for each model in Figure~\ref{fig:deepspeechLayerBreakdown}.
	DeepSpeech includes five multi-batch \texttt{FullyConnected} layers with batch size of 16, and one single-batch 
	\texttt{LSTM} layer.
	As shown in Figure~\ref{fig:deepspeechLayerBreakdown}, the \texttt{LSTM} layer is responsible for more than $70\%$
	of the whole model execution time.
	Please note that we can execute sub-byte models on methods that do not explicitly support them,
	but this leads to no speedup compared to the $W8A8$ models.
	So in this paper, we focus on providing speedup while using sub-byte models.

	Throughout our experiments, we compare \ourmethod{} to nine other methods that are available by default or we managed to add
	to TFLite: (1) \emph{ULPPACK} \cite{won2022ulppack}, which to the best of our knowledge represents the state of the art and is the latest mechanism 
	capable of processing different sub-byte models;
	four of the other eight methods operate on quantized models ($W8A8$\footnote{$Wn$: n bits for the weight values, $Am$: m bits for the activation values; thus, $W8A8$ case represents 8 bits for each of weights and activations.}):
		(2) \emph{Ruy for $W8A8$ models},
		(3) \emph{XNNPack for $W8A8$ models},
		(4) \emph{TFLite default for $W8A8$ models} and
		(5) \emph{GEMMLOWP}.
	The rest of the rivals operate on floating point ($FP32$) models: 
		(6) \emph{Ruy for $FP32$ models},
		(7) \emph{XNNPack for $FP32$ models},
		(8) \emph{TFLite default for $FP32$ models} and
		(9) \emph{Eigen} \cite{website:eigen}.

	We selected \emph{Ruy for the $W8A8$ model} as the baseline in our experiments because this method, along with \emph{XNNPack}, 
	show the best performance among popular publicly available quantization techniques for CPU platforms, but 
	\emph{XNNPack} is not available in several cases and therefore, \emph{Ruy} is the default optimization option on TFLite as well.

	We run our experiments on gem5 cycle-accurate processor simulator \cite{binkert2011gem5,lowe2020gem5}. This allowed us to gather detailed
	reliable execution metrics, such as cache latency, and performance statistics on all the techniques under evaluation.
	We also evaluate the effect of different cache sizes and hierarchies that might not be available in a real device.
	All $FP32$ methods and \emph{ULPPACK} are slower than the main baseline by one or tow orders of magnitude.
	 The only two methods that outperformed the baseline, are \emph{XNNPack for W8A8 models} ($2.4\times$ speedup) and 
	 \ourmethod{} ($3.1\times$ speedup). On average, \ourmethod{} consistently outperforms the other methods when running on models with different sizes.

	 Our measurements on Raspberry Pi 4 (Section~\ref{appendix:real-device-results}) also supports our evaluation.

	Our contributions are as follows:
	\begin{itemize}
		\item We introduce a packing scheme to efficiently pack multiple sub-byte (${<}8$) elements into a single wider (${\geq}8$)
			  value.
			  This packing scheme fully utilizes the memory footprint and bandwidth usage and is designed to reduce extraction
			  overhead.
		\item Then we propose a set of nine different hand-written GEMV assembly kernels utilizing the introduced packing
			  scheme to effectively process different bit-width (1-, 2-, and 4-bits) and add them to TFLite.
		\item We evaluate \ourmethod{} against eight production-ready GEMM/GEMV libraries and the current state-of-the-art,
			  ULPPACK.
			  We also demonstrate the performance improvement of \ourmethod{} on an end-to-end evaluation of Mozilla DeepSpeech.
	\end{itemize}

	The rest of this paper is organized as follows.
	We discuss the related work on quantized model execution on constrained
	devices in Section \ref{RelatedWork}.
	We introduce our method in Section \ref{method} and the results of our
	evaluations in Section \ref{results}. 
	We conclude in Section \ref{conclusion}.

%% file: figures/mozilla-deepspeech-per-layer-breakdown.tex
\begin{figure}[t!]
    \centering
        \begin{tikzpicture}
            \begin{axis}[
                xbar stacked,
                xmin=0,
                xmax=1600,
                width=0.8*\columnwidth,
                height=5cm,
                symbolic y coords={FP32,W8A8,W4A4 (\ourmethod{}),W2A2 (\ourmethod{}),W1A1 (\ourmethod{})},
                tick label style={font=\footnotesize},
                legend style={font=\footnotesize},
                label style={font=\footnotesize},
                xtick={0,200,400,800,1600},
                ytick = data,
                xlabel={Execution Time (milisecond)},
                nodes near coords align={vertical},
                legend pos=north east,
                align={center},
                xmajorgrids,
                yminorgrids = true,
                minor tick num=1,
                legend columns=2,
                legend pos=north east
                ]
                \addplot[Layer1, fill={Layer1}, postaction={pattern=\LayerOnePattern}]
                coordinates {(6.395,FP32) (2.979,W8A8) (2.981,{W4A4 (\ourmethod{})}) (2.987,{W2A2 (\ourmethod{})}) (2.984,{W1A1 (\ourmethod{})})};
                
                \addplot[Layer2, fill={Layer2}, postaction={pattern=\LayerTwoPattern}] 
                coordinates {(27.083,FP32) (11.049,W8A8) (11.052,{W4A4 (\ourmethod{})}) (11.046,{W2A2 (\ourmethod{})}) (11.044,{W1A1 (\ourmethod{})})};
                
                \addplot[Layer3, fill={Layer3}, postaction={pattern=\LayerThreePattern}] 
                coordinates {(26.752,FP32) (10.975,W8A8) (10.983,{W4A4 (\ourmethod{})}) (10.971,{W2A2 (\ourmethod{})}) (10.974,{W1A1 (\ourmethod{})})};
                
                \addplot[LSTM, fill={LSTM}, postaction={pattern=\LayerLSTMPattern}] 
                coordinates {(1511.685,FP32) (358.748,W8A8) (206.71,{W4A4 (\ourmethod{})}) (196.666,{W2A2 (\ourmethod{})}) (136.052,{W1A1 (\ourmethod{})})};
                
                \addplot[Layer5, fill={Layer5}, postaction={pattern=\LayerFourPattern}] 
                coordinates {(26.642,FP32) (11.095,W8A8) (11.106,{W4A4 (\ourmethod{})}) (11.110,{W2A2 (\ourmethod{})}) (11.105,{W1A1 (\ourmethod{})})};
                
                \addplot[Layer6, fill={Layer6}, postaction={pattern=\LayerFivePattern}] 
                coordinates {(0.565,FP32) (0.216,W8A8) (0.219,{W4A4 (\ourmethod{})}) (0.216,{W2A2 (\ourmethod{})}) (0.217,{W1A1 (\ourmethod{})})};

                \legend{
                    Linear 1, Linear 2, Linear 3, LSTM, Linear 5, Linear 6
                }
            \end{axis}
        \end{tikzpicture}
    \vspace{-3mm}
    \caption{
        Mozilla DeepSpeech \cite{website:mozilladeepspeech} per layer execution time breakdown 
        for our W1A1, W2A2, and W4A4 sub-byte quantized, Ruy W8A8 quantized, and Ruy FP32 full precision models.
    }
    \vspace{-4mm}
    \label{fig:deepspeechLayerBreakdown}
\end{figure}
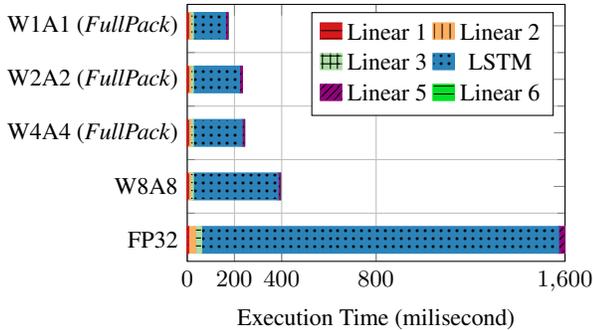

%% file: Sections/relatedwork.tex
\section{Related Works}
\label{RelatedWork}

	Literature has extensively demonstrated different techniques of quantizing a deep learning model to sub-byte
	precision while maintaining or showing negligible drop in accuracy.
	LSQ \cite{esser2019learned} proposes learning quantization parameters. BitPruning
	\cite{nikolic2020bitpruning} presents an approach for learning the bit-width of each cluster of values.
	To make the quantized model even more lightweight, \cite{alom2018effective,courbariaux2015binaryconnect} leveraged binary (-1 or 1) values 
	(weight and/or activations) in different types of DNNs.
	On the other hand, some methods use non-uniform quantization approaches to enhance the performance of their system.
	MAM \cite{chang2021mix} uses a power-of-2 scheme to quantize the model parameters while GOBO \cite{zadeh2020gobo} uses a weight clustering 
	approach, mapping them to only a few unique values.
	While non-uniform quantization approaches report significant accuracy improvements, these methods do not usually
	perform well on the commodity CPUs, or are deemed entirely infeasible due to lack of suitable instrucitons in the CPU ISA.

	Utilizing sub-byte quantized models on mobile devices has also gained attention recently.
	Bison-e \cite{reggiani2022bison} proposed a new hardware extension to RISC-V ISA alongside an approach to pack multiple smaller sub-byte 
	values inside a larger integer. 
	This method utilizes binary segmentation \cite{pan1993binary} to perform GEMM (GEneral Matrix Multiplication) operations.
	ULPPACK \cite{won2022ulppack} utilizes the same method to perform GEMM operations without extending the current ARMv8 ISA.
	ULPPACK offers two packing schemes to pack the sub-byte values into larger (${\geq}16$-bit) integers with an 
	optimized local accumulation to reduce the impact of data extraction overhead.
	Both research directions show satisfying speedups even on end-to-end results, nevertheless each has its own drawbacks;
	Bison-e requires hardware and ISA extensions and thus cannot be used on off-the-shelf general-purpose CPUs, while ULPPACK lacks the ability 
	to fully utilize memory space and bandwidth as well as the full compute capacity of the vector processing units.
	We provide a replacement that solves both above shortcomings. We use only commodity ARMv8 NEON vector instructions, and our data layout 
	mechanism packs the data elements without spacers in between; despite we need to run more instructions to unpack the data after loaded into vector 
	registers, we show that we still get net positive gain since memory access overhead is the dominant factor in many DNN implementations on today 
	CPUs.
	
	In contrast, we use no additional hardware or new instructions.
	Our memory layout and assembly kernels are designed for the Vector Unit so that they fully utilize the memory space
	and execute GEMV operations effectively.

%% file: Sections/method.tex
\section{Method and Implementation}
\label{method}

	\ourmethod{} consists of two main elements:
	\begin{inparaenum}[(1)]
		\item an efficient packing scheme that fully utilizes the memory bandwidth and space while purpose-built for the processing
		steps to be applied in the CPU, and
		\item handwritten ARMv8-A NEON assembly kernels optimized for each bit-width (1-, 2-, and 4-bits)
			to take best advantage from the packing scheme.
	\end{inparaenum}
	Thus it can also be viewed as a storage-processing co-design scheme.
	
	\subsection{Packing Scheme}
	\label{method:packingScheme}

		We design our packing scheme according to the bit-width of weights (activations) and the size
		of the vector register in the VPU (vector-processing unit).
		Our method can be used to pack multiple sub-byte (4-, 2-, or 1-bits) parameters in a single byte on any vector register of
		any size, but we stick to the NEON VPU on ARM CPUs.
		
		The core idea behind our design is minimizing the overhead in the extraction of sub-byte parameters.
		To achieve this, however, we need to know which sub-byte values we should pack together in a specific byte.
		The na\"ive approach packs the adjacent values within an array into a single byte.
		Algorithm \ref{alg:method:naive:W8A8} demonstrates the procedure to process data using na\"ive packing scheme.

		In the na\"ive method, it loads one byte of weight and extracts packed 4-bit values with three shift operations (lines 6-7).
		Then would load two bytes of corresponding activations (lines 6-7) to multiply them by the corresponding weights and
		accumulate the result to the corresponding output value (lines 10-11).

		Although this method fully utilizes memory bandwidth and space, but it is inefficient to use on VPUs, because 
		the extraction overhead dominates.
		Noting that NEON ISA performs logical and arithmetic operations at byte level on 16 bytes in
		parallel using only a single vector instruction, the packing+processing scheme can be co-designed.

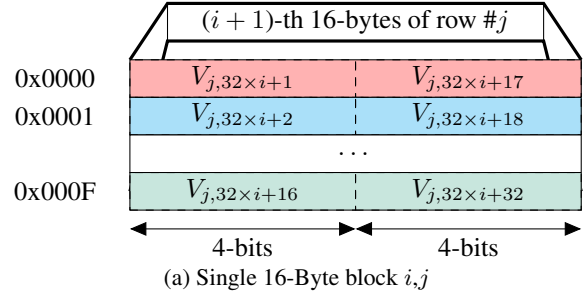
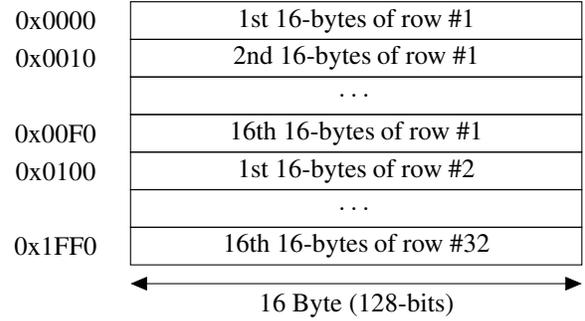
\begin{figure}[t!]
			\centering
			\begin{subfigure}[b]{\columnwidth}
				\centering
				\begin{tikzpicture}[
					Single/.style = {draw, dashed, minimum width=3cm, minimum height=0.5cm, inner sep=0,outer sep=0},
					SingleContainer/.style = {draw, minimum width=6cm, minimum height=0.5cm, inner sep=0,outer sep=0},
					BlockIndicator/.style= {draw, minimum width=5cm, minimum height=0.5cm, inner sep=0,outer sep=0},
					MultipleBytes/.style= {draw, minimum width=6cm, minimum height=0.5cm, inner sep=0,outer sep=0},
					Color1/.style = {fill=red!30},
					Color2/.style = {fill=cyan!30},
					Color3/.style = {fill=SeaGreen!30},
					ColorWhite/.style = {fill=white!30},
					]
					
					\node[BlockIndicator] (TIJ) at (0, 0.25) {($i+1$)-th 16-bytes of row \#$j$};
	
					\draw[black, very thick,] (-3,-0.25) -- (-2.5,0.5) -- (2.5,0.5) -- (3,-0.25);
					\draw[black, very thick,] (-3,-2) -- (-2.5,0) -- (2.5,0) -- (3,-2);
					
					\node[SingleContainer, Color1] (CIJ1A17) at (0, -0.5) {};
					\node[Single] (NIJ1) at (-1.5,-0.5) {$V_{j,32 \times i + 1}$};
					\node[Single] (NIJ17) at (1.5,-0.5) {$V_{j,32 \times i + 17}$};
					\node[draw=none] (LIJ1A17) at (-4,-0.5) {0x0000};
	
					\node[SingleContainer, Color2] (CIJ2A18) at (0, -1) {};
					\node[Single] (NIJ2) at (-1.5,-1) {$V_{j,32 \times i + 2}$};
					\node[Single] (NIJ18) at (1.5,-1) {$V_{j,32 \times i + 18}$};
					\node[draw=none] (LIJ2A18) at (-4,-1) {0x0001};
	
					\node[MultipleBytes, ColorWhite] (CIJ2A18) at (0, -1.5) {$\dots$};
	
					\node[SingleContainer, Color3] (CIJ16A32) at (0, -2) {};
					\node[Single] (NIJ16) at (-1.5,-2) {$V_{j,32 \times i + 16}$};
					\node[Single] (NIJ32) at (1.5,-2) {$V_{j,32 \times i + 32}$};
					\node[draw=none] (LIJ16A32) at (-4,-2) {0x000F};
	
					\draw[>=triangle 45, <->] (-3,-2.5) -- (0,-2.5) node[midway,below] {4-bits};
					\draw[>=triangle 45, <->] (0,-2.5) -- (3,-2.5)  node[midway,below] {4-bits};
				\end{tikzpicture}
				\vspace{-2mm}
				\caption{
					\centering
					Single 16-Byte block $i$,$j$
				}
				\vspace{-3mm}
				\label{fig:method:packingScheme:4bitScheme-block}
			\end{subfigure}
			\begin{subfigure}[b]{\columnwidth}
				\vspace*{0.5cm}
				\centering
				\begin{tikzpicture}[
					BlockIndicator/.style= {draw, minimum width=6cm, minimum height=0.5cm, inner sep=0,outer sep=0},
					]
					
					\node[BlockIndicator] (NI1J1) at (0, 0) {1st 16-bytes of row \#1};
					\node[draw=none] (LI1J1) at (-4,0) {0x0000};

					\node[BlockIndicator] (NI1J2) at (0, -0.5) {2nd 16-bytes of row \#1};
					\node[draw=none] (LI1J2) at (-4,-0.5) {0x0010};

					\node[BlockIndicator] (NI1J3T15) at (0, -1) {$\dots$};
					
					\node[BlockIndicator] (NI1J16) at (0, -1.5) {16th 16-bytes of row \#1};
					\node[draw=none] (LI1J16) at (-4,-1.5) {0x00F0};

					\node[BlockIndicator] (NI2J1) at (0, -2) {1st 16-bytes of row \#2};
					\node[draw=none] (LI2J1) at (-4,-2) {0x0100};

					\node[BlockIndicator] (NI2T32J15) at (0, -2.5) {$\dots$};

					\node[BlockIndicator] (NI32J16) at (0, -3) {16th 16-bytes of row \#32};
					\node[draw=none] (LI32J16) at (-4,-3) {0x1FF0};
	
					\draw[>=triangle 45, <->] (-3,-3.5) -- (3,-3.5) node[midway,below] {16 Byte (128-bits)};
				\end{tikzpicture}
				\vspace{-2mm}
				\caption{
					\centering
					In-memory placment of the example
				}
				\label{fig:method:packingScheme:4bitScheme-all}
			\end{subfigure}
			\vspace{-6mm}
			\caption{
				Proposed packing scheme for an example matrix of size ${32{\times}16}$ with 4-bit values.
			}
			\vspace{-5mm}
			\label{fig:method:packingScheme:4bitScheme}
		\end{figure}

		For 4-bit quantization, our idea is to pack every two 4-bit elements with stride 16, into a single byte; then, store
		every 16 of such packs adjacently on consecutive bytes; then again, put every 16 of such 16-byte packs (from the same rows of the matrix)
		consecutively. This is repeated to fully cover one set of rows of the matrix, and is then repeated again 
		for all other sets of rows of the matrix.
		Figure \ref{fig:method:packingScheme:4bitScheme} demonstrates our proposal for 4-bit values in a ${32{\times}16}$ matrix.

		\begin{algorithm}[b!]
			\begin{small}
				\caption{Na\"ive method for the W4A8 model using FMA}
				\label{alg:method:naive:W8A8}
				\begin{algorithmic}[1]
					\REQUIRE weights and $k$-dimensional activation as $W$ and $A$ with $8$ and $8$ bit width
					\ENSURE $z$-dimensional output $O$
					\STATE{$i \gets 0$}
					\FOR{$i < z$}
						\STATE{$j \gets 0$}
						\STATE{$O[i] \gets 0$}
						\FOR{$j < k$}
							\vspace*{5pt}
							\STATE{$W_0 \gets (W[i] \gg 4) \ll 4$}
							\STATE{$W_1 \gets W[i] \ll 4$}
							\vspace*{5pt}
							\STATE{$A_0 \gets A[i]$}
							\STATE{$A_1 \gets A[i + 1]$}
							\vspace*{5pt}
							\STATE{$O[i] \gets$ FMA($W_0$ , $A_0$ , $O[i]$)}
							\STATE{$O[i] \gets$ FMA($W_1$ , $A_1$ , $O[i]$)}
							\vspace*{5pt}
							\STATE{$j \gets j + 2$}
						\ENDFOR
						\STATE{$O[i] \gets$ ElementWiseAdd($V_2$)}
						\STATE{$i \gets i + 1$}
					\ENDFOR
					\STATE{\textbf{return} $O$}
				\end{algorithmic}
			\end{small}
		\end{algorithm}
		
		Then at processing time using vector instructions, the above arrangement allows to load 16-bytes of data at once into the vector
		registers, and extract every 16 originally-subsequent elements by simple vector-shift operations.
		Note that the 16-bytes data is read only once from memory, but it contains 32 elements of the original 4-bit values;
		compare that with ULPPack storage where spacer bits are put between elements, and hence, same 16-byte read from memory yields
		smaller amount of useful data. 
		
		Thus, VPU can extract values from 1 to 16 with two shifts (one logical shift to the left and one arithmetic shift to 
		the right to do the sign-extension), and then the 16th to 32nd values with one arithmetic shift to the right to do the 
		sign-extension.
		The reason behind the two required shifts for extraction of values 1-16 is that by applying two shifts, we can
		mask and sign-extend the values but with only masking, we can not sign-extend values.
		On the other hand, these two shifts can be performed in place. Therefore, there is no need for another VPU register.

		Our packing scheme can pack 1-bit and 2-bit parameters as well.
		For 1-bit values, we pack eight 1-bit values with stride 16 in a single byte, and thus 128 original 1-bit values are loaded from
		memory into a vector register by a single 16-byte vector load instruction. 
		Similarly, for 2-bit values, there would be four values in a single byte and 64 values loaded from memory into a vector register.
		We can selectively apply our packing scheme to weights or activations or both.
		Also, one can utilize any VPU with any bit width by extending the above scheme.
		Obviously, with larger vector registers, we can fit more values inside a single vector. However, for scalable vector schemes such as 
		ARM SVE and RISC-V RVV, our processing scheme needs adjustments since the processor vector length is not known statically; 
		this remains as part of our future work. 

	\subsection{GEMV Kernels}
	\label{method:mainMethod}

		To perform matrix multiplication on weights and activations, we provide a series of kernel functions.
		Each function is speciallized to process a specific model type. The supported types are $W8A4$, $W4A8$, $W4A4$,
		$W2A8$, $W8A2$, $W2A2$, $W1A8$, $W8A1$, and $W1A1$.
		
		\input{figures/method-w4a8-single-iterations.tex}
		
		Algorithm~\ref{alg:method:mainMethod:W4A8} shows how we process the $W4A8$ model using Fused
		Multiply and Add (FMA) instruction.
		In each function, we load a vector-size block of data (weights or activations) into a vector register and if needed, based on the 
		quantization bit-width of the model, we extract/load the values into one or more vector registers (lines 6-11), and then we multiply 
		the extracted weights and activations and accumulate the resulting products in a vector register (lines 12-13).
		Figure~\ref{fig:method:mainMethod:W4A8} tries to illustrate the processing of a block of weights and the corresponding activations 
		for the model.
		As depicted, the weights are loaded from memory into a vector register, and then demultiplexed into two vector registers 
		($V_0$ and $V_1$).
		For extracting the weights $W_1$ to $W_{16}$ of this block, we need two shift operations: one Logical Shift Left (LSL) for masking and 
		one Arithmetic Shift Right (ASR) for sign extension. For extracting the other 16 weights ($W_{17}$ to $W_{32}$) from the original vector 
		register, only one Arithmetic Shift Right (ASR) operation is needed for sign extension.
		After preparing the weights, we load 32 corresponding activations into two vectors, $V_2$ and $V_3$, and then
		process the dot-product of 32 weights and activations in those vector registers.

		\begin{algorithm}[t]
			\begin{small}
			\caption{Proposed method for the W4A8 model using FMA and Vector ISA}
			\label{alg:method:mainMethod:W4A8}
			\begin{algorithmic}[1]
				\REQUIRE Packed weights and $k$-dimensional activation as $W$ and $A$ with $4$ and $8$ bit width
				\ENSURE $z$-dimensional output $O$
				\STATE{$i \gets 0$}
				\FOR{$i < z$}
					\STATE{$j \gets 0$}
					\STATE{$V_4 \gets 0$}
					\FOR{$j < k$}
						\vspace*{5pt}
						\STATE{$V_0 \gets$ load 16 bytes of $W$ and increment}
						\STATE{$V_1 \gets$ ArithmeticShiftRight($V_0$ , $4$)}
						\STATE{$V_0 \gets$ LogicalShiftLeft($V_0$ , $4$)}
						\STATE{$V_0 \gets$ ArithmeticShiftRight($V_0$ , $4$)}
						\vspace*{5pt}
						\STATE{$V_2 \gets$ load 16 bytes of $A$ and increment}
						\STATE{$V_3 \gets$ load 16 bytes of $A$ and increment}
						\vspace*{5pt}
						\STATE{$V_4 \gets$ FMA($V_0$ , $V_2$, $V_4$)} 
						\STATE{$V_4 \gets$ FMA($V_1$ , $V_3$, $V_4$)}
						\STATE{$j \gets j + 32$}
					\ENDFOR
					\STATE{$O[i] \gets$ ElementWiseAdd($V_4$)}
					\STATE{$i \gets i + 1$}
				\ENDFOR
				\STATE{\textbf{return} $O$}
			\end{algorithmic}
		\end{small}
		\end{algorithm}

%% file: figures/method-w4a8-single-iterations.tex
\begin{figure}[t!]
    \centering
    \begin{adjustbox}{max width=\columnwidth}
        \begin{tikzpicture}[
            Operator/.style = {draw, minimum width=0.5cm, minimum height=0.5cm},
            Node/.style = {draw, minimum width=1cm, minimum height=1cm},
            Single/.style = {draw, dashed, minimum width=1cm, minimum height=1cm},
            SingleContainer/.style = {draw, minimum width=2cm, minimum height=1cm},
            BlockIndicator/.style= {draw, minimum width=5cm, minimum height=1cm},
            MultipleBytes/.style= {draw, minimum width=2cm, minimum height=1cm},
            FullBlock/.style= {draw, minimum width=8cm, minimum height=1cm},
            Color1/.style = {fill=red!30},
            Color2/.style = {fill=cyan!30},
            Color3/.style = {fill=SeaGreen!30},
            ColorWhite/.style = {fill=white!30},
            ]
            \draw[>=triangle 45, ->] (3,1.5) -- (3,0.5) node[midway,right] {Load 16 Bytes of weights from memory};

            \node[SingleContainer, Color1] (C1A17) at (0,0) {};
            \node[Single] (N17) at (-0.5,0) {$W_{17}$};
            \node[Single] (N1) at (0.5,0) {$W_1$};

            \node[SingleContainer, Color2] (C2A18) at (2,0) {};
            \node[Single] (N18) at (1.5,0) {$W_{18}$};
            \node[Single] (N2) at (2.5,0) {$W_2$};
            
            \node[SingleContainer] (CM3A19T15A31) at (4,0) {$\dots$};

            \node[SingleContainer, Color3] (C16A32) at (6,0) {};
            \node[Single] (N32) at (5.5,0) {$W_{32}$};
            \node[Single] (N16) at (6.5,0) {$W_{16}$};

            \node[SingleContainer, Color1] (C1A17) at (0 - 4,0 - 1.5) {};
            \node[Single] (N17) at (-0.5 - 4,0 - 1.5) {$W_1$};
            \node[Single] (N1) at (0.5 - 4,0 - 1.5) {0x0};

            \node[SingleContainer, Color2] (C2A18) at (2 - 4,0 - 1.5) {};
            \node[Single] (N18) at (1.5 - 4,0 - 1.5) {$W_2$};
            \node[Single] (N2) at (2.5 - 4,0 - 1.5) {0x0};
            
            \node[SingleContainer] (CM3A19T15A31) at (4 - 4,0 - 1.5) {$\dots$};

            \node[SingleContainer, Color3] (C16A32) at (6 - 4,0 - 1.5) {};
            \node[Single] (N32) at (5.5 - 4,0 - 1.5) {$W_{16}$};
            \node[Single] (N16) at (6.5 - 4,0 - 1.5) {0x0};

            \draw[>=triangle 45, ->] (-1,0) -- (-2.5,0) 
                node[near end,above] {LSL} -- (-2.5,-1);

            \node[SingleContainer, Color1] (C1A17) at (0 + 4,0 - 3.5) {};
            \node[Single] (N17) at (-0.5 + 4,0 - 3.5) {Sign};
            \node[Single] (N1) at (0.5 + 4,0 - 3.5) {$W_{17}$};

            \node[SingleContainer, Color2] (C2A18) at (2 + 4,0 - 3.5) {};
            \node[Single] (N18) at (1.5 + 4,0 - 3.5) {Sign};
            \node[Single] (N2) at (2.5 + 4,0 - 3.5) {$W_{18}$};
            
            \node[SingleContainer] (CM3A19T15A31) at (4 + 4,0 - 3.5) {$\dots$};

            \node[SingleContainer, Color3] (C16A32) at (6 + 4,0 - 3.5) {};
            \node[Single] (N32) at (5.5 + 4,0 - 3.5) {Sign};
            \node[Single] (N16) at (6.5 + 4,0 - 3.5) {$W_{32}$};

            \draw[>=triangle 45, ->] (7,0) -- (8.5,0) 
                node[near end,above] {ASR} -- (8.5,-3);

            \node[SingleContainer, Color1] (C1A17) at (0 - 4,0 - 3.5) {};
            \node[Single] (N17) at (-0.5 - 4,0 - 3.5) {Sign};
            \node[Single] (N1) at (0.5 - 4,0 - 3.5) {$W_1$};

            \node[SingleContainer, Color2] (C2A18) at (2 - 4,0 - 3.5) {};
            \node[Single] (N18) at (1.5 - 4,0 - 3.5) {Sign};
            \node[Single] (N2) at (2.5 - 4,0 - 3.5) {$W_2$};
            
            \node[SingleContainer] (CM3A19T15A31) at (4 - 4,0 - 3.5) {$\dots$};

            \node[SingleContainer, Color3] (C16A32) at (6 - 4,0 - 3.5) {};
            \node[Single] (N32) at (5.5 - 4,0 - 3.5) {Sign};
            \node[Single] (N16) at (6.5 - 4,0 - 3.5) {$W_{16}$};

            \draw[>=triangle 45, ->] (-2.5,-2) -- (-2.5,-3) node[midway,left] {ASR};

            \draw[>=triangle 45, <->] (-5,-4.2) -- (3,-4.2) 
                                    node[midway,below] {$V_0$};
            \draw[>=triangle 45, <->] (3,-4.2) -- (11,-4.2) 
                                    node[midway,below] {$V_1$};

            \node[FullBlock] (A17T32) at (7,-5.5) {Activations $17$ To $32$};

            \node[FullBlock] (A1T16) at (-1,-5.5) {Activations $1$ To $16$};

            \draw[>=triangle 45, <->] (-5,-6.2) -- (3,-6.2) 
                                    node[midway,below] {$V_2$};
            \draw[>=triangle 45, <->] (3,-6.2) -- (11,-6.2) 
                                    node[midway,below] {$V_3$};

            \node[Operator] (Mul1T16) at (7,-7.25) {$\times$};
            \node[Operator] (Mul17T32) at (-1,-7.25) {$\times$};

            \draw[>=triangle 45, ->] (1,-4.2) -- (1,-7.25) -- (-0.75,-7.25);
            \draw[>=triangle 45, ->] (-3,-6.2) -- (-3,-7.25) -- (-1.25,-7.25);
            
            \draw[>=triangle 45, ->] (5,-4.2) -- (5,-7.25) -- (6.75,-7.25);
            \draw[>=triangle 45, ->] (9,-6.2) -- (9,-7.25) -- (7.25,-7.25);

            \node[Operator] (Add1T16B17T32) at (3, -7.75) {$+$};
            
            \draw[>=triangle 45, ->] (Mul1T16)  -- (7,-7.75) -- (Add1T16B17T32);
            \draw[>=triangle 45, ->] (Mul17T32) -- (-1,-7.75) -- (Add1T16B17T32);

            \node[Node] (O1T32) at (3, -9) {Output};

            \draw[>=triangle 45, ->] (Add1T16B17T32) -- (O1T32);
        \end{tikzpicture}
    \end{adjustbox}
    \vspace{-6mm}
    \caption{
        Processing a block of weights (16 Bytes) with the proposed method for the W4A8 model.
        $V_0$, $V_1$, $V_2$, and $V_3$ are four sample vector registers and the output is a scalar accumulated in the $V_4$ register.
    }
    \vspace{-4mm}
    \label{fig:method:mainMethod:W4A8}
\end{figure}
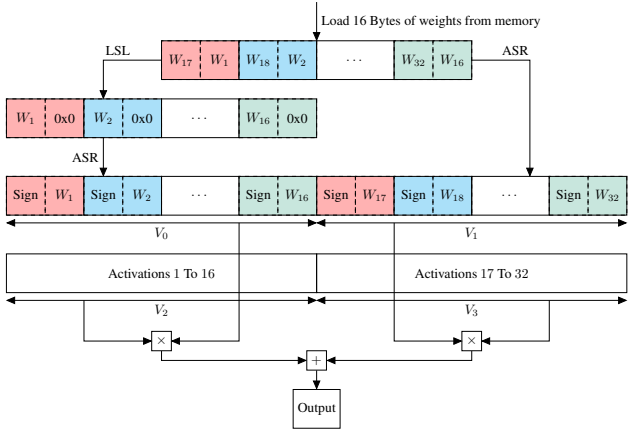

%% file: Sections/evaluation.tex
\section{Evaluation}
\label{results}

	We implemented and integrated our method into TFLite and compared it to eight production-ready
	libraries, mostly written in assembly by industry experts, and available in TFLite, as well as \ullpack{}.

	\begin{table}[!b]
		\vspace{-7mm}
		\caption{\emph{gem5} simulation setup}
		\label{tab:evaluation:gem5-spec}
		\begin{small}
			\centering
			\begin{tabular}{ll}
				\hline
				
				CPU Type & modified ex5\_big\\
				
				Architecture & ARMv8-A \\
				Micro-Architecture & \scalebox{0.9}{Single core @ 2.45GHz (max freq.)}\\

				L1 Cache   (per core) & 128KB I\$ + 128KB D\$ \\
				
				L2 Cache$^{\times}$ (Shared) & 2 MB\\

				L3 Cache (Shared) & 8 MB (where employed)\\
				
				RAM & 4GB (LPDDR3x @ 1600MHz)\\
				\hline
				
			\end{tabular}
			\raggedright\newline
			$^{\times}$: Size may alter or get removed based on the experiment.
		\end{small}
	\end{table}

	\subsection{Experiments Setup}\label{results:setup}

		We evaluated all nine methods on the cycle-accurate \emph{gem5} simulator \cite{binkert2011gem5,lowe2020gem5} 
		(Table~\ref{tab:evaluation:gem5-spec}).
		Except in \S\ref{results:LLCacheVariable} where we evaluate different L2 cache sizes and also add an L3 cache, 
		the default configurations for the rest of the experiments include a 2MB L2 cache as the last level cache.

		Additionally, we evaluated \ourmethod{} on Raspberry Pi 4 (\S~\ref{appendix:real-device-results}), and compared it with the other methods on FullyConnected layers of eleven well-known convolutional neural networks.
	
		We employed TFLite benchmarking tool \cite{website:tfliteBenchmarkTool} which we built using \texttt{-c opt} Bazel
		flag that enables \texttt{-O3} flag on compile-time. 
		This tool allows us to easily select the method we want to run with run-time flags.
		However, \gemmlowp{} and \eigen{} needed a compile-time flag to be activated to replace the default execution path.

		We evaluated our proposal against the following methods.
		
		\textbf{Ruy for W8A8 models.}
		(\ruyint{}):
		\ruy{} is the default method in TFLite when the caching is enabled. This method is developed by Google and is
		the fastest among all rivals, except \xnnpack{}.
		
		\textbf{XNNPack.}
		(\xnnpackint{}):
		This method by Google and Facebook is often the fastest method in TFLite, but it requires heavy
		preprocessing and does not support all operations of TFLite; this causes slowdown when moving data between
		supported and not supported operators. 
		In addition, our focus is on the efficacy of the packing scheme. 
		Thus, the other ISA-specific instructions, e.g., prefetching are out of the scope of this work. 
		\xnnpackint{} is written in assembly.
		
		\input{figures/performance_comparison.tex}
		
		\textbf{TFLite default for W8A8 models.}
		(\tfliteint{}):
		This method is the default method when the caching is disabled and is written in C/C++ with compiler intrinsics.
		
		\textbf{GEMMLOWP.}
		(\gemmlowpint{}):
		Another library in TFLite for GEMM and GEMV operations that is not available by default but can be
		embedded into the binaries using a compile-time flag. Unlike \ruy{} and \xnnpack{}, this library only supports $W8A8$ models
		and does not support 32-bit floating point ($FP32$) models. \gemmlowp{} is also handwritten in assembly.
		
		\textbf{Ruy for FP32 models.}
		(\ruyfp{}):
		This method is also from the \ruy{} library but for processing $FP32$ models.
		
		\textbf{XNNPack for FP32 models.}
		(\xnnpackfp{}):
		This method is also from the \xnnpack{} library but for processing $FP32$ models.
		
		\textbf{TFLite default for FP32 models.}
		(\tflitefp{}):
		This method is the default execution mechanism that TFLite employs for processing $FP32$ models while caching is not enabled.
		
		\textbf{Eigen.}
		(\eigenfp{}):
		This method only supports $FP32$ models and like \gemmlowp{}, is not available by default, but can be added to the
		binary using a compile-time flag.
		
		\textbf{ULPPACK.} 
		\footnote{\ullpack{} does not have an open-source code base. We contacted the authors and they kindly
		sent us the codes; we cordially acknowledge and appreciate that. 
		We then ported it to TFLite ourselves for comparisons, so beware of any potential deficiencies inadvertently introduced.}
		(\ullpackwa{W3A3}, \ullpackwa{W2A2}, \ullpackwa{W1A1}):
		\ullpack{} supports sub-byte models with different bit-widths for activations and weights independently.
		However, we only selected models with the same bit-width for activations and weights for brevity.
		Also note that according to the authors \cite{won2022ulppack}, \ullpack{} does not gain speedup with $W4A4$,
		$W5A5$, $W6A6$, and $W7A7$ models.
		Further note that \ullpack{} only implements GEMM and does not have any GEMV-specific kernel, so
		in each inference experiment, we pass \ullpack{} an input with 8 batches with the same values; we call this reduced version, \ullpack{}$^-$.

		\textbf{The Baseline.}
		Although \xnnpackint{} was often the fastest among the rivals, but since it is not available for some cases of our studies or
		degrades performance for the same,
		we chose \ruyint{} as the baseline, and normalize all results against it.

	\subsection{Performance Comparison}\label{results:performance}

		\input{figures/different-quantization-schemes.tex}

		We first study the speedup on different input and output sizes of a \texttt{FullyConnected} layer.
		Figure~\ref{fig:results:performance:mainFig} depicts the results.
		Except \ourmethod{} for $W4A8$ and \xnnpackint{}, all other methods are slower than the baseline (\ruyint{}).
		\xnnpackint{}, in contrast to our method, gains more speedup for smaller models while our method outperforms it
		for larger models.
		This is mainly because \ourmethod{} uses less memory bandwidth which is more visible on larger inputs.
		We will discuss this more in \S\ref{results:WeightsVsActivaitons:LLCacheBehavior}.
		As we can see in Figure~\ref{fig:results:performance:mainFig}, on average, our method for $W4A8$ can reach a performance
		gain of $2.44\times$.

	\subsection{What to Quantize? Weights, Activations, or Both?}\label{results:WeightsVsActivaitons}

		Another aspect is the effect of different quantization schemes.
		In \S\ref{results:performance}, we evaluated our method on the $W4A8$ model which only utilizes quantization on 
		the weights of the model, however we can also quantize only the activations ($W8A4$), or weights and activations together
		($W4A4$).

		\input{figures/llc.tex}

		\input{figures/bitwidth.tex}

		Different assembly kernels to process the GEMV for each model are required.
		To run $W8A4$ models, we apply our packing scheme only on activations while for $W4A4$ models, we apply our 
		packing scheme on both activations and weights.
		
		Figure~\ref{fig:results:WeightsVsActivaitons:mainFig} illustrates the result of running our method on $W8A4$
		and $W4A4$ models.
		Applying sub-byte quantization on the weights will cause a speedup of 2.44$\times$ while 
		applying the same quantization only on activations improves the performance by 1.92$\times$.
		However, if we apply sub-byte quantization on both weights and activations, the performance will improve by 
		2.48$\times$ which is only about 1.02$\times$ faster than the model that only has sub-byte weights.

		\input{figures/mozilla-deepspeech-arch.tex}

		The reason behind this is that total size of weight elements in GEMV operations is in general bigger than the activations.
		Consequently, when we quantize the weights, memory bandwidth usage drops significantly compared to quantizing the
		activations.

		By taking a closer look at Figures~\ref{fig:results:performance:ourW4A8},
		\ref{fig:results:WeightsVsActivaitons:ourW8A4}, and \ref{fig:results:WeightsVsActivaitons:ourW4A4}, a diagonal boundary from 
		the top right to the bottom left of each table is noticeable and demonstrates higher speedups compared to other cases. 
		From this boundary to the left (before the boundary), we mostly observe a reduction in speedup. To the right of the boundary, however, 
		we observe higher speedup compared to the left-hand side, but it saturates: almost no change with respect to the modification 
		of input/output sizes.
		This can be justified by the effect of caches which we discuss in \S\ref{results:WeightsVsActivaitons:LLCacheBehavior}.

		\subsubsection{Last-Level Cache Behavior}\label{results:WeightsVsActivaitons:LLCacheBehavior}

			To investigate the mentioned observation, we evaluate the Last Level Cache (LLC) behavior throughout the execution of each model.
			Figure~\ref{fig:results:WeightsVsActivaitons:LLCacheBehavior:mainFig} demonstrates the LLC behavior on models of different sizes.
			Using sub-byte weights (Figure~\ref{fig:results:WeightsVsActivaitons:LLCacheBehavior:Accesses:ourW4A8}), the LLC accesses are
			reduced by $50\%$ for larger models, while sub-byte activations barely reduce LLC accesses for these models
			(Figure~\ref{fig:results:WeightsVsActivaitons:LLCacheBehavior:Accesses:ourW8A4} for $W8A4$ and
			Figure~\ref{fig:results:WeightsVsActivaitons:LLCacheBehavior:Accesses:ourW4A4} for $W4A4$).

			However, if we take a closer look at the nodes to the right of the diagonal of $W4A8$ models in
			Figure~\ref{fig:results:WeightsVsActivaitons:LLCacheBehavior:mainFig}, it illustrates that accesses of our method 
			(\ref{fig:results:WeightsVsActivaitons:LLCacheBehavior:Accesses:ourW4A8}) are 50\% fewer than the baseline,
			and misses (\ref{fig:results:WeightsVsActivaitons:LLCacheBehavior:Misses:ourW4A8}) are about 90\% fewer than the
			baseline.
			This causes the miss rate (\ref{fig:results:WeightsVsActivaitons:LLCacheBehavior:MissRate:ourW4A8}) to be 70-80\% 
			lower than the baseline.
			Looking at these sizes and the L2 cache size in Table~\ref{tab:evaluation:gem5-spec}, we find out
			that in these sizes, our weight matrix fits in the L2 cache but the W8A8 weight matrix does not.
			This causes the baseline to suffer from $\sim$99\% L2 cache miss rate.
			Figure~(\ref{fig:results:WeightsVsActivaitons:LLCacheBehavior:Latency:ourW4A8}) depicts LLC
			cache miss latency. 
			It shows that our method reduces the cache miss latency by 80-90\% on these models.

			Furthermore, after the diagonal boundary, the LLC accesses
			(\ref{fig:results:WeightsVsActivaitons:LLCacheBehavior:Accesses:ourW4A8})
			and misses (\ref{fig:results:WeightsVsActivaitons:LLCacheBehavior:Misses:ourW4A8}) are both reduced by 50\%, and thus the same
			miss rate (\ref{fig:results:WeightsVsActivaitons:LLCacheBehavior:MissRate:ourW4A8}) as the baseline. 
			This demonstrates the case where our method takes best advantage from its lower memory bandwidth usage and reduces the LLC 
			miss latency (\ref{fig:results:WeightsVsActivaitons:LLCacheBehavior:Latency:ourW4A8}) by $\sim$50\%.

			Note how $W8A4$ model has almost same number of accesses, misses (and thus miss rate) and LLC miss latency as the baseline 
			at IO sizes to the right of the diagonal boundary (\ref{fig:results:WeightsVsActivaitons:LLCacheBehavior:Accesses:ourW8A4},
			\ref{fig:results:WeightsVsActivaitons:LLCacheBehavior:Misses:ourW8A4},
			\ref{fig:results:WeightsVsActivaitons:LLCacheBehavior:MissRate:ourW8A4}, and
			\ref{fig:results:WeightsVsActivaitons:LLCacheBehavior:Latency:ourW8A4}). This confirms that activation-quantization is not
			as effective as weight-quantization here.

	\subsection{Different Sizes of Last-Level Cache}
	\label{results:LLCacheVariable}

		In the previous Section, we showed that the formation of a maximum-speedup boundary obtained by our method is an effect of 
		the Last-Level Cache capacity; when even the packed data does not fit the LLC, we start to lose some speedup while still 
		performing better than \ruyint{}.
		We further evaluated the above behavior on our $W4A4$ model vs. baseline under various cache sizes and cache hierarchies.
		Figure~\ref{fig:results:LLCacheVariable:mainFig} presents the results;
		at higher LLC sizes or when an L3 cache is introduced, the maximum-speedup boundary moves to the higher IO sizes. 
		This further confirms and also quntifies the LLC effect on our obtained speedups. 
		Even with L2 and L3 caches removed (\ref{fig:results:LLCacheVariable:RemovedL2}), the same above effect is observed
		but at smaller sizes since now L1 size is the limit.
		Note that the inference latency differs when cache size and structure changes; the above figures only depict the \emph{speedup}
		vs. baseline in each case.

	\subsection{What If We Use Fewer Bits?}\label{results:DifferentBitwidths}
		In the previous experiments, we only evaluated our method on 4-bit quantized models.
		Here, we evaluate the narrower bit widths for weights and/or activations.
		Figure~\ref{fig:results:DifferentBitwidths:mainFig} shows speedups and instructions count of our method on $W2A2$ and
		$W1A1$ models w.r.t. our method on $W4A4$.
		Using fewer bits expands the maximum-speedup boundary region and also improves the obtainable speedup beyond that boundary.
		However, if we observe figures \ref{fig:results:DifferentBitwidths:inst:ourW2A2} and
		\ref{fig:results:DifferentBitwidths:inst:ourW1A1} we can see that compared to $W4A4$, \ourmethod{} for $W2A2$ models
		uses negligible fewer instructions but $W1A1$ models requires more instructions: 
		$1.03\times$ for $W2A2$, and $0.8\times$ for $W1A1$ models compared to $W4A4$ models.
		Such behavior leads to a higher speedup on larger models compared to $W4A4$:
		$1.23\times$ for $W2A2$, and $1.17\times$ for $W1A1$ models.

		\input{figures/end-to-end-results.tex}

	\subsection{End-To-End Performance}\label{results:endToEnd}

		For the end-to-end performance, we evaluate all the methods on Mozilla DeepSpeech \cite{website:mozilladeepspeech}.
		This model contains five multi-batch Fully Connected layers with 16 batches and one multi-batch LSTM layer with 16 batches
		which is unrolled to 16 consecutive single-batch LSTM layers.
		The model architecture is shown in Figure~\ref{fig:results:endToEnd:MozillaDeepSpeech}.
		Only LSTM layers are single-batch. These single-batch layers are the layers that utilize GEMV operations;  other layers use GEMM
		operations. However, as depicted in Figure~\ref{fig:deepspeechLayerBreakdown}, the LSTM layer consumes
		more than 70\% of the whole inference time.
		Since our algorithm is for the GEMV operations, we apply it only on the LSTM layer and we use \ruyint{} (the baseline) for the 
		GEMM operations (multi-batch layers).

		Figure~\ref{fig:results:endToEnd:deepspeechLayerBreakdown} illustrates end-to-end breakdown of per-layer execution time of the
		DeepSpeech model for each method, extracted with the TFLite benchmarking tool per operation profiling.
		Regarding the total execution time, we observe that \ourmethod{} for all three models outperforms all the
		others despite the fact that all of our improvement comes only from the LSTM layer.
		Our method can achieve an end-to-end speedup of 1.56-2.11$\times$ and 1.23-1.66$\times$ compared to
		the best rivals, namely (\ruyint{}) and \xnnpackint{} respectively.

		For more results of the evaluation of all methods on real-world models on a real device, please refer to the section
		\ref{appendix:real-device-results} of the appendix.

		\subsection{On-Device Performance}\label{results:onDevice}
		For evaluation on the real devices, we selected Fully Connected layers of eleven well-known Convolutional Neural Networks (CNNs), namely DenseNet201, EfficientNetV2L, InceptionV3, InceptionResNetV2, MobileNetV2, NASNetLarge, RegNetY320, ResNet152, ResNet152V2, VGG19, and Xception. These layers in CNNs utilize GEMV, which is the main focus of this paper, while the other layers, including convolutional layers, are implemented with GEMM operations.
		We executed each of them using TFLite benchmarking tool, for 10 warmup iterations and 100 main iterations on Raspberry Pi 4
		(Table~\ref{tab:appendix:real-devices}) and averaged over the main iterations for the results.
		Figure~\ref{fig:appendix:endToEnd:CNNsFCs-rapsberry-pi-4} (in Appendix) demonstrates the speedup of each method over the baseline, \ruyint{}. The results further support our evaluation on \emph{gem5} as we achieve on average $1.2{\times}$, $1.5{\times}$ and $1.43{\times}$, and up to $1.38{\times}$, $1.69{\times}$ and $1.62{\times}$ speedup over the main baseline for \texttt{W1A1}, \texttt{W2A2} and \texttt{W4A4}, respectively, while outperforming the other rivals.

		\begin{table}[b!]
			\vspace{-7mm}
			\captionof{table}{Raspberry Pi 4 Model B Specifications}
			\label{tab:appendix:real-devices}
			\centering
			\begin{tabular}{ll}
				\hline
				CPU Type & Broadcom BCM2711\\
				Architecture & ARMv8-A\\
				Micro-Architecture & \scalebox{0.9}{4$\times$ (Cortex-A72) core }\\
								   & @ 2.45GHz (max freq.)\\
				L1 Cache (per core) & 32KB I + 32KB D\\
				L2 Cache (Shared) & 1 MB\\
				RAM & 4GB (LPDDR4 @ 2400MHz)\\
				\hline
			\end{tabular}
		\end{table}

%% file: figures/different-quantization-schemes.tex
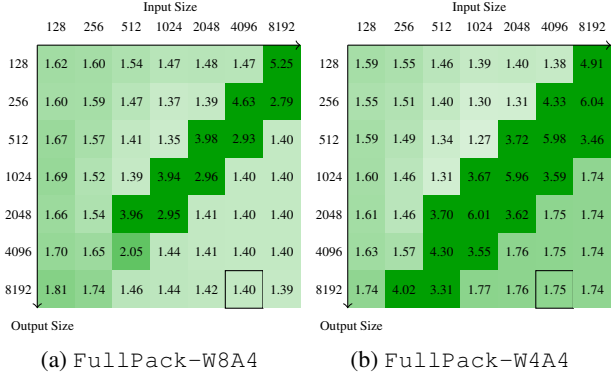
\begin{figure}[t!]
    \newcommand{\CreateGradientColorCell}[3]{%
        \ifdim #1 pt > \MidNumber pt
            \pgfmathsetmacro{\PercentColor}{max(min(100.0*(#1 - \MidNumber)/(\MaxNumber-\MidNumber),100.0),0.00)} %
            \node[draw=#3, Node, fill=\MaxColor!\PercentColor!\MidColor] at #2 {\scalebox{0.5}{#1}};
        \else
            \pgfmathsetmacro{\PercentColor}{max(min(100.0*(\MidNumber - #1)/(\MidNumber-\MinNumber),100.0),0.00)} %
            \node[draw=#3, Node, fill=\MinColor!\PercentColor!\MidColor] at #2 {\scalebox{0.5}{#1}};
        \fi
    }
    \newcommand{\CreateGradientColorCellWithName}[5]{%
        \ifdim #1 pt > \MidNumber pt
            \pgfmathsetmacro{\PercentColor}{max(min(100.0*(#1 - \MidNumber)/(\MaxNumber-\MidNumber),100.0),0.00)} %
            \node[draw=#3, #5, fill=\MaxColor!\PercentColor!\MidColor] at #2 {\setstretch{0.8}\scalebox{0.6}{#4}\\\scalebox{0.6}{#1}};
        \else
            \pgfmathsetmacro{\PercentColor}{max(min(100.0*(\MidNumber - #1)/(\MidNumber-\MinNumber),100.0),0.00)} %
            \node[draw=#3, #5, fill=\MinColor!\PercentColor!\MidColor] at #2 {\setstretch{0.8}\scalebox{0.6}{#4}\\\scalebox{0.6}{#1}};
        \fi
    }
    \newcommand{\CreateLogTwoGradientColorCell}[3]{%
        \ifdim #1 pt > \MidNumber pt
            \pgfmathsetmacro{\PercentColor}{max(min(100.0*(#1 - \MidNumber)/(\MaxNumber-\MidNumber),100.0),0.00)} %
            \node[draw=#3, Node, fill=\MaxColor!\PercentColor!\MidColor] at #2 {\scalebox{0.5}{$2^{#1}$}};
        \else
            \pgfmathsetmacro{\PercentColor}{max(min(100.0*(\MidNumber - #1)/(\MidNumber-\MinNumber),100.0),0.00)} %
            \node[draw=#3, Node, fill=\MinColor!\PercentColor!\MidColor] at #2 {\scalebox{0.5}{$2^{#1}$}};
        \fi
    }
    \centering
    \newcommand*{\HighlighColor}{black}%
    \begin{subfigure}[b]{0.49\linewidth}
        \centering
        \newcommand*{\MinNumber}{1.2}%
        \newcommand*{\MidNumber}{1.0} %
        \newcommand*{\MaxNumber}{2.7}%
        \newcommand*{\MinColor}{red}%
        \newcommand*{\MidColor}{white} %
        \newcommand*{\MaxColor}{maxGreen}%
        \newcommand{\baseX}{0}%
        \newcommand{\baseY}{0}%
        \newcommand*{\MinWidth}{0.5}%
        \begin{tikzpicture}[
            Node/.style = {minimum width=\MinWidth cm, minimum height=\MinWidth cm, inner sep=0,outer sep=0},
        ]
            
            \node[Node] at (\baseX + 0 * \MinWidth,\baseY + \MinWidth) {\scalebox{\MinWidth}{128}};
            \node[Node] at (\baseX + 1 * \MinWidth,\baseY + \MinWidth) {\scalebox{\MinWidth}{256}};
            \node[Node] at (\baseX + 2 * \MinWidth,\baseY + \MinWidth) {\scalebox{\MinWidth}{512}};
            \node[Node] at (\baseX + 3 * \MinWidth,\baseY + \MinWidth) {\scalebox{\MinWidth}{1024}};
            \node[Node] at (\baseX + 4 * \MinWidth,\baseY + \MinWidth) {\scalebox{\MinWidth}{2048}};
            \node[Node] at (\baseX + 5 * \MinWidth,\baseY + \MinWidth) {\scalebox{\MinWidth}{4096}};
            \node[Node] at (\baseX + 6 * \MinWidth,\baseY + \MinWidth) {\scalebox{\MinWidth}{8192}};
            
            \node[Node] at (\baseX - \MinWidth,\baseY - 0 * \MinWidth) {\scalebox{\MinWidth}{128}};
            \node[Node] at (\baseX - \MinWidth,\baseY - 1 * \MinWidth) {\scalebox{\MinWidth}{256}};
            \node[Node] at (\baseX - \MinWidth,\baseY - 2 * \MinWidth) {\scalebox{\MinWidth}{512}};
            \node[Node] at (\baseX - \MinWidth,\baseY - 3 * \MinWidth) {\scalebox{\MinWidth}{1024}};
            \node[Node] at (\baseX - \MinWidth,\baseY - 4 * \MinWidth) {\scalebox{\MinWidth}{2048}};
            \node[Node] at (\baseX - \MinWidth,\baseY - 5 * \MinWidth) {\scalebox{\MinWidth}{4096}};
            \node[Node] at (\baseX - \MinWidth,\baseY - 6 * \MinWidth) {\scalebox{\MinWidth}{8192}};

            \renewcommand{\baseY}{-1 * 0 * \MinWidth}%
            \CreateGradientColorCell{1.62}{(\baseX + 0 * \MinWidth,\baseY - 0.0)}{none}
            \CreateGradientColorCell{1.60}{(\baseX + 1 * \MinWidth,\baseY - 0.0)}{none}
            \CreateGradientColorCell{1.54}{(\baseX + 2 * \MinWidth,\baseY - 0.0)}{none}
            \CreateGradientColorCell{1.47}{(\baseX + 3 * \MinWidth,\baseY - 0.0)}{none}
            \CreateGradientColorCell{1.48}{(\baseX + 4 * \MinWidth,\baseY - 0.0)}{none}
            \CreateGradientColorCell{1.47}{(\baseX + 5 * \MinWidth,\baseY - 0.0)}{none}
            \CreateGradientColorCell{5.25}{(\baseX + 6 * \MinWidth,\baseY - 0.0)}{none}
            \renewcommand{\baseY}{-1 * 1 * \MinWidth}%
            \CreateGradientColorCell{1.60}{(\baseX + 0 * \MinWidth,\baseY - 0.0)}{none}
            \CreateGradientColorCell{1.59}{(\baseX + 1 * \MinWidth,\baseY - 0.0)}{none}
            \CreateGradientColorCell{1.47}{(\baseX + 2 * \MinWidth,\baseY - 0.0)}{none}
            \CreateGradientColorCell{1.37}{(\baseX + 3 * \MinWidth,\baseY - 0.0)}{none}
            \CreateGradientColorCell{1.39}{(\baseX + 4 * \MinWidth,\baseY - 0.0)}{none}
            \CreateGradientColorCell{4.63}{(\baseX + 5 * \MinWidth,\baseY - 0.0)}{none}
            \CreateGradientColorCell{2.79}{(\baseX + 6 * \MinWidth,\baseY - 0.0)}{none}
            \renewcommand{\baseY}{-1 * 2 * \MinWidth}%
            \CreateGradientColorCell{1.67}{(\baseX + 0 * \MinWidth,\baseY - 0.0)}{none}
            \CreateGradientColorCell{1.57}{(\baseX + 1 * \MinWidth,\baseY - 0.0)}{none}
            \CreateGradientColorCell{1.41}{(\baseX + 2 * \MinWidth,\baseY - 0.0)}{none}
            \CreateGradientColorCell{1.35}{(\baseX + 3 * \MinWidth,\baseY - 0.0)}{none}
            \CreateGradientColorCell{3.98}{(\baseX + 4 * \MinWidth,\baseY - 0.0)}{none}
            \CreateGradientColorCell{2.93}{(\baseX + 5 * \MinWidth,\baseY - 0.0)}{none}
            \CreateGradientColorCell{1.40}{(\baseX + 6 * \MinWidth,\baseY - 0.0)}{none}
            \renewcommand{\baseY}{-1 * 3 * \MinWidth}%
            \CreateGradientColorCell{1.69}{(\baseX + 0 * \MinWidth,\baseY - 0.0)}{none}
            \CreateGradientColorCell{1.52}{(\baseX + 1 * \MinWidth,\baseY - 0.0)}{none}
            \CreateGradientColorCell{1.39}{(\baseX + 2 * \MinWidth,\baseY - 0.0)}{none}
            \CreateGradientColorCell{3.94}{(\baseX + 3 * \MinWidth,\baseY - 0.0)}{none}
            \CreateGradientColorCell{2.96}{(\baseX + 4 * \MinWidth,\baseY - 0.0)}{none}
            \CreateGradientColorCell{1.40}{(\baseX + 5 * \MinWidth,\baseY - 0.0)}{none}
            \CreateGradientColorCell{1.40}{(\baseX + 6 * \MinWidth,\baseY - 0.0)}{none}
            \renewcommand{\baseY}{-1 * 4 * \MinWidth}%
            \CreateGradientColorCell{1.66}{(\baseX + 0 * \MinWidth,\baseY - 0.0)}{none}
            \CreateGradientColorCell{1.54}{(\baseX + 1 * \MinWidth,\baseY - 0.0)}{none}
            \CreateGradientColorCell{3.96}{(\baseX + 2 * \MinWidth,\baseY - 0.0)}{none}
            \CreateGradientColorCell{2.95}{(\baseX + 3 * \MinWidth,\baseY - 0.0)}{none}
            \CreateGradientColorCell{1.41}{(\baseX + 4 * \MinWidth,\baseY - 0.0)}{none}
            \CreateGradientColorCell{1.40}{(\baseX + 5 * \MinWidth,\baseY - 0.0)}{none}
            \CreateGradientColorCell{1.40}{(\baseX + 6 * \MinWidth,\baseY - 0.0)}{none}
            \renewcommand{\baseY}{-1 * 5 * \MinWidth}%
            \CreateGradientColorCell{1.70}{(\baseX + 0 * \MinWidth,\baseY - 0.0)}{none}
            \CreateGradientColorCell{1.65}{(\baseX + 1 * \MinWidth,\baseY - 0.0)}{none}
            \CreateGradientColorCell{2.05}{(\baseX + 2 * \MinWidth,\baseY - 0.0)}{none}
            \CreateGradientColorCell{1.44}{(\baseX + 3 * \MinWidth,\baseY - 0.0)}{none}
            \CreateGradientColorCell{1.41}{(\baseX + 4 * \MinWidth,\baseY - 0.0)}{none}
            \CreateGradientColorCell{1.40}{(\baseX + 5 * \MinWidth,\baseY - 0.0)}{none}
            \CreateGradientColorCell{1.40}{(\baseX + 6 * \MinWidth,\baseY - 0.0)}{none}
            \renewcommand{\baseY}{-1 * 6 * \MinWidth}%
            \CreateGradientColorCell{1.81}{(\baseX + 0 * \MinWidth,\baseY - 0.0)}{none}
            \CreateGradientColorCell{1.74}{(\baseX + 1 * \MinWidth,\baseY - 0.0)}{none}
            \CreateGradientColorCell{1.46}{(\baseX + 2 * \MinWidth,\baseY - 0.0)}{none}
            \CreateGradientColorCell{1.44}{(\baseX + 3 * \MinWidth,\baseY - 0.0)}{none}
            \CreateGradientColorCell{1.42}{(\baseX + 4 * \MinWidth,\baseY - 0.0)}{none}
            \CreateGradientColorCell{1.40}{(\baseX + 5 * \MinWidth,\baseY - 0.0)}{\HighlighColor}
            \CreateGradientColorCell{1.39}{(\baseX + 6 * \MinWidth,\baseY - 0.0)}{none}

            \draw[->] (-1 * \MinWidth / 2,\MinWidth / 2) -- (\MinWidth * 6.5,\MinWidth / 2);
            \node[minimum height=\MinWidth cm, inner sep=0,outer sep=0] at (1.5,0.75) {\scalebox{\MinWidth}{ Input Size}};

            \draw[->] (-1 * \MinWidth / 2,\MinWidth / 2) -- (-1 * \MinWidth / 2,-1 * \MinWidth * 6.5);
            \node[minimum height=\MinWidth cm, inner sep=0,outer sep=0] at (-0.2,-3.5) {\scalebox{\MinWidth}{ Output Size}};
        \end{tikzpicture}
        \vspace{-6mm}
        \caption{
            \centering
            \ourmethodWA{W8A4}
        }
        \label{fig:results:WeightsVsActivaitons:ourW8A4}
    \end{subfigure}
    \begin{subfigure}[b]{0.49\linewidth}
        \centering
        \newcommand*{\MinNumber}{1.2}%
        \newcommand*{\MidNumber}{1.0} %
        \newcommand*{\MaxNumber}{2.7}%
        \newcommand*{\MinColor}{red}%
        \newcommand*{\MidColor}{white} %
        \newcommand*{\MaxColor}{maxGreen}%
        \newcommand{\baseX}{0}%
        \newcommand{\baseY}{0}%
        \newcommand*{\MinWidth}{0.5}%
        \begin{tikzpicture}[
            Node/.style = {minimum width=\MinWidth cm, minimum height=\MinWidth cm, inner sep=0,outer sep=0},
        ]

            \node[Node] at (\baseX + 0 * \MinWidth,\baseY + \MinWidth) {\scalebox{\MinWidth}{128}};
            \node[Node] at (\baseX + 1 * \MinWidth,\baseY + \MinWidth) {\scalebox{\MinWidth}{256}};
            \node[Node] at (\baseX + 2 * \MinWidth,\baseY + \MinWidth) {\scalebox{\MinWidth}{512}};
            \node[Node] at (\baseX + 3 * \MinWidth,\baseY + \MinWidth) {\scalebox{\MinWidth}{1024}};
            \node[Node] at (\baseX + 4 * \MinWidth,\baseY + \MinWidth) {\scalebox{\MinWidth}{2048}};
            \node[Node] at (\baseX + 5 * \MinWidth,\baseY + \MinWidth) {\scalebox{\MinWidth}{4096}};
            \node[Node] at (\baseX + 6 * \MinWidth,\baseY + \MinWidth) {\scalebox{\MinWidth}{8192}};
            
            \node[Node] at (\baseX - \MinWidth,\baseY - 0 * \MinWidth) {\scalebox{\MinWidth}{128}};
            \node[Node] at (\baseX - \MinWidth,\baseY - 1 * \MinWidth) {\scalebox{\MinWidth}{256}};
            \node[Node] at (\baseX - \MinWidth,\baseY - 2 * \MinWidth) {\scalebox{\MinWidth}{512}};
            \node[Node] at (\baseX - \MinWidth,\baseY - 3 * \MinWidth) {\scalebox{\MinWidth}{1024}};
            \node[Node] at (\baseX - \MinWidth,\baseY - 4 * \MinWidth) {\scalebox{\MinWidth}{2048}};
            \node[Node] at (\baseX - \MinWidth,\baseY - 5 * \MinWidth) {\scalebox{\MinWidth}{4096}};
            \node[Node] at (\baseX - \MinWidth,\baseY - 6 * \MinWidth) {\scalebox{\MinWidth}{8192}};

            \renewcommand{\baseY}{-1 * 0 * \MinWidth}%
            \CreateGradientColorCell{1.59}{(\baseX + 0 * \MinWidth,\baseY - 0.0)}{none}
            \CreateGradientColorCell{1.55}{(\baseX + 1 * \MinWidth,\baseY - 0.0)}{none}
            \CreateGradientColorCell{1.46}{(\baseX + 2 * \MinWidth,\baseY - 0.0)}{none}
            \CreateGradientColorCell{1.39}{(\baseX + 3 * \MinWidth,\baseY - 0.0)}{none}
            \CreateGradientColorCell{1.40}{(\baseX + 4 * \MinWidth,\baseY - 0.0)}{none}
            \CreateGradientColorCell{1.38}{(\baseX + 5 * \MinWidth,\baseY - 0.0)}{none}
            \CreateGradientColorCell{4.91}{(\baseX + 6 * \MinWidth,\baseY - 0.0)}{none}
            \renewcommand{\baseY}{-1 * 1 * \MinWidth}%
            \CreateGradientColorCell{1.55}{(\baseX + 0 * \MinWidth,\baseY - 0.0)}{none}
            \CreateGradientColorCell{1.51}{(\baseX + 1 * \MinWidth,\baseY - 0.0)}{none}
            \CreateGradientColorCell{1.40}{(\baseX + 2 * \MinWidth,\baseY - 0.0)}{none}
            \CreateGradientColorCell{1.30}{(\baseX + 3 * \MinWidth,\baseY - 0.0)}{none}
            \CreateGradientColorCell{1.31}{(\baseX + 4 * \MinWidth,\baseY - 0.0)}{none}
            \CreateGradientColorCell{4.33}{(\baseX + 5 * \MinWidth,\baseY - 0.0)}{none}
            \CreateGradientColorCell{6.04}{(\baseX + 6 * \MinWidth,\baseY - 0.0)}{none}
            \renewcommand{\baseY}{-1 * 2 * \MinWidth}%
            \CreateGradientColorCell{1.59}{(\baseX + 0 * \MinWidth,\baseY - 0.0)}{none}
            \CreateGradientColorCell{1.49}{(\baseX + 1 * \MinWidth,\baseY - 0.0)}{none}
            \CreateGradientColorCell{1.34}{(\baseX + 2 * \MinWidth,\baseY - 0.0)}{none}
            \CreateGradientColorCell{1.27}{(\baseX + 3 * \MinWidth,\baseY - 0.0)}{none}
            \CreateGradientColorCell{3.72}{(\baseX + 4 * \MinWidth,\baseY - 0.0)}{none}
            \CreateGradientColorCell{5.98}{(\baseX + 5 * \MinWidth,\baseY - 0.0)}{none}
            \CreateGradientColorCell{3.46}{(\baseX + 6 * \MinWidth,\baseY - 0.0)}{none}
            \renewcommand{\baseY}{-1 * 3 * \MinWidth}%
            \CreateGradientColorCell{1.60}{(\baseX + 0 * \MinWidth,\baseY - 0.0)}{none}
            \CreateGradientColorCell{1.46}{(\baseX + 1 * \MinWidth,\baseY - 0.0)}{none}
            \CreateGradientColorCell{1.31}{(\baseX + 2 * \MinWidth,\baseY - 0.0)}{none}
            \CreateGradientColorCell{3.67}{(\baseX + 3 * \MinWidth,\baseY - 0.0)}{none}
            \CreateGradientColorCell{5.96}{(\baseX + 4 * \MinWidth,\baseY - 0.0)}{none}
            \CreateGradientColorCell{3.59}{(\baseX + 5 * \MinWidth,\baseY - 0.0)}{none}
            \CreateGradientColorCell{1.74}{(\baseX + 6 * \MinWidth,\baseY - 0.0)}{none}
            \renewcommand{\baseY}{-1 * 4 * \MinWidth}%
            \CreateGradientColorCell{1.61}{(\baseX + 0 * \MinWidth,\baseY - 0.0)}{none}
            \CreateGradientColorCell{1.46}{(\baseX + 1 * \MinWidth,\baseY - 0.0)}{none}
            \CreateGradientColorCell{3.70}{(\baseX + 2 * \MinWidth,\baseY - 0.0)}{none}
            \CreateGradientColorCell{6.01}{(\baseX + 3 * \MinWidth,\baseY - 0.0)}{none}
            \CreateGradientColorCell{3.62}{(\baseX + 4 * \MinWidth,\baseY - 0.0)}{none}
            \CreateGradientColorCell{1.75}{(\baseX + 5 * \MinWidth,\baseY - 0.0)}{none}
            \CreateGradientColorCell{1.74}{(\baseX + 6 * \MinWidth,\baseY - 0.0)}{none}
            \renewcommand{\baseY}{-1 * 5 * \MinWidth}%
            \CreateGradientColorCell{1.63}{(\baseX + 0 * \MinWidth,\baseY - 0.0)}{none}
            \CreateGradientColorCell{1.57}{(\baseX + 1 * \MinWidth,\baseY - 0.0)}{none}
            \CreateGradientColorCell{4.30}{(\baseX + 2 * \MinWidth,\baseY - 0.0)}{none}
            \CreateGradientColorCell{3.55}{(\baseX + 3 * \MinWidth,\baseY - 0.0)}{none}
            \CreateGradientColorCell{1.76}{(\baseX + 4 * \MinWidth,\baseY - 0.0)}{none}
            \CreateGradientColorCell{1.75}{(\baseX + 5 * \MinWidth,\baseY - 0.0)}{none}
            \CreateGradientColorCell{1.74}{(\baseX + 6 * \MinWidth,\baseY - 0.0)}{none}
            \renewcommand{\baseY}{-1 * 6 * \MinWidth}%
            \CreateGradientColorCell{1.74}{(\baseX + 0 * \MinWidth,\baseY - 0.0)}{none}
            \CreateGradientColorCell{4.02}{(\baseX + 1 * \MinWidth,\baseY - 0.0)}{none}
            \CreateGradientColorCell{3.31}{(\baseX + 2 * \MinWidth,\baseY - 0.0)}{none}
            \CreateGradientColorCell{1.77}{(\baseX + 3 * \MinWidth,\baseY - 0.0)}{none}
            \CreateGradientColorCell{1.76}{(\baseX + 4 * \MinWidth,\baseY - 0.0)}{none}
            \CreateGradientColorCell{1.75}{(\baseX + 5 * \MinWidth,\baseY - 0.0)}{\HighlighColor}
            \CreateGradientColorCell{1.74}{(\baseX + 6 * \MinWidth,\baseY - 0.0)}{none}

            \draw[->] (-1 * \MinWidth / 2,\MinWidth / 2) -- (\MinWidth * 6.5,\MinWidth / 2);
            \node[minimum height=\MinWidth cm, inner sep=0,outer sep=0] at (1.5,0.75) {\scalebox{\MinWidth}{Input Size}};

            \draw[->] (-1 * \MinWidth / 2,\MinWidth / 2) -- (-1 * \MinWidth / 2,-1 * \MinWidth * 6.5);
            \node[minimum height=\MinWidth cm, inner sep=0,outer sep=0] at (-0.2,-3.5) {\scalebox{\MinWidth}{Output Size}};
        \end{tikzpicture}
        \vspace{-6mm}
        \caption{
            \centering
            \ourmethodWA{W4A4}
        }
        \label{fig:results:WeightsVsActivaitons:ourW4A4}
    \end{subfigure}
    \centering
    \vspace{-3mm}
    \caption{
        \centering
        The effect of quantizing weights and/or activations on the performance (Speedup) of our method on different
        sizes.
    }
    \vspace{-4mm}
    \label{fig:results:WeightsVsActivaitons:mainFig}
\end{figure}

%% file: figures/bitwidth.tex
\begin{figure}[t!]
    \newcommand{\CreateGradientColorCell}[3]{%
        \ifdim #1 pt > \MidNumber pt
            \pgfmathsetmacro{\PercentColor}{max(min(100.0*(#1 - \MidNumber)/(\MaxNumber-\MidNumber),100.0),0.00)} %
            \node[draw=#3, Node, fill=\MaxColor!\PercentColor!\MidColor] at #2 {\scalebox{\nodeScale}{#1}};
        \else
            \pgfmathsetmacro{\PercentColor}{max(min(100.0*(\MidNumber - #1)/(\MidNumber-\MinNumber),100.0),0.00)} %
            \node[draw=#3, Node, fill=\MinColor!\PercentColor!\MidColor] at #2 {\scalebox{\nodeScale}{#1}};
        \fi
    }
    \newcommand{\CreateGradientColorCellSaturate}[3]{%
        \ifdim #1 pt > \MaxNumber pt
            \node[draw=#3, Node, fill=\MaxColor] at #2 {\scalebox{\nodeScale}{$\geq$\MaxNumber}};
        \else
            \ifdim #1 pt = \MaxNumber pt
                \node[draw=#3, Node, fill=\MaxColor] at #2 {\scalebox{\nodeScale}{$\geq$\MaxNumber}};
            \else
                \ifdim #1 pt > \MidNumber pt
                    \pgfmathsetmacro{\PercentColor}{max(min(100.0*(#1 - \MidNumber)/(\MaxNumber-\MidNumber),100.0),0.00)} %
                    \node[draw=#3, Node, fill=\MaxColor!\PercentColor!\MidColor] at #2 {\scalebox{\nodeScale}{#1}};
                \else
                    \pgfmathsetmacro{\PercentColor}{max(min(100.0*(\MidNumber - #1)/(\MidNumber-\MinNumber),100.0),0.00)} %
                    \node[draw=#3, Node, fill=\MinColor!\PercentColor!\MidColor] at #2 {\scalebox{\nodeScale}{#1}};
                \fi
            \fi
        \fi
    }
    \newcommand{\CreateGradientColorCellWithName}[5]{%
        \ifdim #1 pt > \MidNumber pt
            \pgfmathsetmacro{\PercentColor}{max(min(100.0*(#1 - \MidNumber)/(\MaxNumber-\MidNumber),100.0),0.00)} %
            \node[draw=#3, #5, fill=\MaxColor!\PercentColor!\MidColor] at #2 {\setstretch{0.8}\scalebox{0.6}{#4}\\\scalebox{0.6}{#1}};
        \else
            \pgfmathsetmacro{\PercentColor}{max(min(100.0*(\MidNumber - #1)/(\MidNumber-\MinNumber),100.0),0.00)} %
            \node[draw=#3, #5, fill=\MinColor!\PercentColor!\MidColor] at #2 {\setstretch{0.8}\scalebox{0.6}{#4}\\\scalebox{0.6}{#1}};
        \fi
    }
    \newcommand{\CreateLogTwoGradientColorCell}[3]{%
        \ifdim #1 pt > \MidNumber pt
            \pgfmathsetmacro{\PercentColor}{max(min(100.0*(#1 - \MidNumber)/(\MaxNumber-\MidNumber),100.0),0.00)} %
            \node[draw=#3, Node, fill=\MaxColor!\PercentColor!\MidColor] at #2 {\scalebox{\nodeScale}{$2^{#1}$}};
        \else
            \pgfmathsetmacro{\PercentColor}{max(min(100.0*(\MidNumber - #1)/(\MidNumber-\MinNumber),100.0),0.00)} %
            \node[draw=#3, Node, fill=\MinColor!\PercentColor!\MidColor] at #2 {\scalebox{\nodeScale}{$2^{#1}$}};
        \fi
    }
    \centering
    \newcommand*{\HighlighColor}{black}%
    \newcommand{\baseX}{0}%
    \newcommand{\baseY}{0}%
    \newcommand*{\MinWidth}{0.5}%
    \newcommand*{\nodeScale}{0.5}%
    \begin{subfigure}[b]{0.49\linewidth}
        \centering
        \newcommand*{\MinColor}{red}%
        \newcommand*{\MidColor}{white} %
        \newcommand*{\MaxColor}{maxGreen}%
        \newcommand*{\MinNumber}{0.00}%
        \newcommand*{\MidNumber}{1.00} %
        \newcommand*{\MaxNumber}{2.00}%
        \begin{tikzpicture}[
            Node/.style = {minimum width=\MinWidth cm, minimum height=\MinWidth cm, inner sep=0,outer sep=0},
        ]

            \node[Node] at (\baseX + 0 * \MinWidth,\baseY + \MinWidth) {\scalebox{\MinWidth}{128}};
            \node[Node] at (\baseX + 1 * \MinWidth,\baseY + \MinWidth) {\scalebox{\MinWidth}{256}};
            \node[Node] at (\baseX + 2 * \MinWidth,\baseY + \MinWidth) {\scalebox{\MinWidth}{512}};
            \node[Node] at (\baseX + 3 * \MinWidth,\baseY + \MinWidth) {\scalebox{\MinWidth}{1024}};
            \node[Node] at (\baseX + 4 * \MinWidth,\baseY + \MinWidth) {\scalebox{\MinWidth}{2048}};
            \node[Node] at (\baseX + 5 * \MinWidth,\baseY + \MinWidth) {\scalebox{\MinWidth}{4096}};
            \node[Node] at (\baseX + 6 * \MinWidth,\baseY + \MinWidth) {\scalebox{\MinWidth}{8192}};
            
            \node[Node] at (\baseX - \MinWidth,\baseY - 0 * \MinWidth) {\scalebox{\MinWidth}{128}};
            \node[Node] at (\baseX - \MinWidth,\baseY - 1 * \MinWidth) {\scalebox{\MinWidth}{256}};
            \node[Node] at (\baseX - \MinWidth,\baseY - 2 * \MinWidth) {\scalebox{\MinWidth}{512}};
            \node[Node] at (\baseX - \MinWidth,\baseY - 3 * \MinWidth) {\scalebox{\MinWidth}{1024}};
            \node[Node] at (\baseX - \MinWidth,\baseY - 4 * \MinWidth) {\scalebox{\MinWidth}{2048}};
            \node[Node] at (\baseX - \MinWidth,\baseY - 5 * \MinWidth) {\scalebox{\MinWidth}{4096}};
            \node[Node] at (\baseX - \MinWidth,\baseY - 6 * \MinWidth) {\scalebox{\MinWidth}{8192}};

            \renewcommand{\baseY}{-1 * 0 * \MinWidth}%
            \CreateGradientColorCell{1.01}{(\baseX + 0 * \MinWidth,\baseY - 0.0)}{none}
            \CreateGradientColorCell{1.02}{(\baseX + 1 * \MinWidth,\baseY - 0.0)}{none}
            \CreateGradientColorCell{1.03}{(\baseX + 2 * \MinWidth,\baseY - 0.0)}{none}
            \CreateGradientColorCell{1.03}{(\baseX + 3 * \MinWidth,\baseY - 0.0)}{none}
            \CreateGradientColorCell{1.05}{(\baseX + 4 * \MinWidth,\baseY - 0.0)}{none}
            \CreateGradientColorCell{1.05}{(\baseX + 5 * \MinWidth,\baseY - 0.0)}{none}
            \CreateGradientColorCell{1.05}{(\baseX + 6 * \MinWidth,\baseY - 0.0)}{none}
            \renewcommand{\baseY}{-1 * 1 * \MinWidth}%
            \CreateGradientColorCell{1.02}{(\baseX + 0 * \MinWidth,\baseY - 0.0)}{none}
            \CreateGradientColorCell{1.02}{(\baseX + 1 * \MinWidth,\baseY - 0.0)}{none}
            \CreateGradientColorCell{1.04}{(\baseX + 2 * \MinWidth,\baseY - 0.0)}{none}
            \CreateGradientColorCell{1.05}{(\baseX + 3 * \MinWidth,\baseY - 0.0)}{none}
            \CreateGradientColorCell{1.05}{(\baseX + 4 * \MinWidth,\baseY - 0.0)}{none}
            \CreateGradientColorCell{1.05}{(\baseX + 5 * \MinWidth,\baseY - 0.0)}{none}
            \CreateGradientColorCell{1.05}{(\baseX + 6 * \MinWidth,\baseY - 0.0)}{none}
            \renewcommand{\baseY}{-1 * 2 * \MinWidth}%
            \CreateGradientColorCell{1.02}{(\baseX + 0 * \MinWidth,\baseY - 0.0)}{none}
            \CreateGradientColorCell{1.03}{(\baseX + 1 * \MinWidth,\baseY - 0.0)}{none}
            \CreateGradientColorCell{1.07}{(\baseX + 2 * \MinWidth,\baseY - 0.0)}{none}
            \CreateGradientColorCell{1.05}{(\baseX + 3 * \MinWidth,\baseY - 0.0)}{none}
            \CreateGradientColorCell{1.05}{(\baseX + 4 * \MinWidth,\baseY - 0.0)}{none}
            \CreateGradientColorCell{1.05}{(\baseX + 5 * \MinWidth,\baseY - 0.0)}{none}
            \CreateGradientColorCell{1.85}{(\baseX + 6 * \MinWidth,\baseY - 0.0)}{none}
            \renewcommand{\baseY}{-1 * 3 * \MinWidth}%
            \CreateGradientColorCell{1.03}{(\baseX + 0 * \MinWidth,\baseY - 0.0)}{none}
            \CreateGradientColorCell{1.05}{(\baseX + 1 * \MinWidth,\baseY - 0.0)}{none}
            \CreateGradientColorCell{1.07}{(\baseX + 2 * \MinWidth,\baseY - 0.0)}{none}
            \CreateGradientColorCell{1.05}{(\baseX + 3 * \MinWidth,\baseY - 0.0)}{none}
            \CreateGradientColorCell{1.05}{(\baseX + 4 * \MinWidth,\baseY - 0.0)}{none}
            \CreateGradientColorCell{1.78}{(\baseX + 5 * \MinWidth,\baseY - 0.0)}{none}
            \CreateGradientColorCell{2.24}{(\baseX + 6 * \MinWidth,\baseY - 0.0)}{none}
            \renewcommand{\baseY}{-1 * 4 * \MinWidth}%
            \CreateGradientColorCell{1.05}{(\baseX + 0 * \MinWidth,\baseY - 0.0)}{none}
            \CreateGradientColorCell{1.04}{(\baseX + 1 * \MinWidth,\baseY - 0.0)}{none}
            \CreateGradientColorCell{1.07}{(\baseX + 2 * \MinWidth,\baseY - 0.0)}{none}
            \CreateGradientColorCell{1.05}{(\baseX + 3 * \MinWidth,\baseY - 0.0)}{none}
            \CreateGradientColorCell{1.76}{(\baseX + 4 * \MinWidth,\baseY - 0.0)}{none}
            \CreateGradientColorCell{2.32}{(\baseX + 5 * \MinWidth,\baseY - 0.0)}{none}
            \CreateGradientColorCell{1.04}{(\baseX + 6 * \MinWidth,\baseY - 0.0)}{none}
            \renewcommand{\baseY}{-1 * 5 * \MinWidth}%
            \CreateGradientColorCell{1.04}{(\baseX + 0 * \MinWidth,\baseY - 0.0)}{none}
            \CreateGradientColorCell{1.05}{(\baseX + 1 * \MinWidth,\baseY - 0.0)}{none}
            \CreateGradientColorCell{1.07}{(\baseX + 2 * \MinWidth,\baseY - 0.0)}{none}
            \CreateGradientColorCell{1.82}{(\baseX + 3 * \MinWidth,\baseY - 0.0)}{none}
            \CreateGradientColorCell{2.22}{(\baseX + 4 * \MinWidth,\baseY - 0.0)}{none}
            \CreateGradientColorCell{1.05}{(\baseX + 5 * \MinWidth,\baseY - 0.0)}{none}
            \CreateGradientColorCell{1.05}{(\baseX + 6 * \MinWidth,\baseY - 0.0)}{none}
            \renewcommand{\baseY}{-1 * 6 * \MinWidth}%
            \CreateGradientColorCell{1.05}{(\baseX + 0 * \MinWidth,\baseY - 0.0)}{none}
            \CreateGradientColorCell{1.05}{(\baseX + 1 * \MinWidth,\baseY - 0.0)}{none}
            \CreateGradientColorCell{2.02}{(\baseX + 2 * \MinWidth,\baseY - 0.0)}{none}
            \CreateGradientColorCell{2.04}{(\baseX + 3 * \MinWidth,\baseY - 0.0)}{none}
            \CreateGradientColorCell{1.05}{(\baseX + 4 * \MinWidth,\baseY - 0.0)}{none}
            \CreateGradientColorCell{1.05}{(\baseX + 5 * \MinWidth,\baseY - 0.0)}{\HighlighColor}
            \CreateGradientColorCell{1.05}{(\baseX + 6 * \MinWidth,\baseY - 0.0)}{none}

            \draw[->] (-1 * \MinWidth / 2,\MinWidth / 2) -- (\MinWidth * 6.5,\MinWidth / 2);
            \node[minimum height=\MinWidth cm, inner sep=0,outer sep=0] at (3 * \MinWidth,1.5 * \MinWidth) {\scalebox{\MinWidth}{ Input Size}};

            \draw[->] (-1 * \MinWidth / 2,\MinWidth / 2) -- (-1 * \MinWidth / 2,-1 * \MinWidth * 6.5);
            \node[minimum height=\MinWidth cm, inner sep=0,outer sep=0] at (-0.4 * \MinWidth,-7 * \MinWidth) {\scalebox{\MinWidth}{ Output Size}};
        \end{tikzpicture}
        \vspace{-3mm}
        \caption{
            \centering
            Execution Time of \ourmethodWA{W2A2}
        }
        \label{fig:results:DifferentBitwidths:ourW2A2}
    \end{subfigure}
    \begin{subfigure}[b]{0.49\linewidth}
        \centering
        \newcommand*{\MinColor}{red}%
        \newcommand*{\MidColor}{white} %
        \newcommand*{\MaxColor}{maxGreen}%
        \newcommand*{\MinNumber}{0.00}%
        \newcommand*{\MidNumber}{1.00} %
        \newcommand*{\MaxNumber}{2.00}%
        \begin{tikzpicture}[
            Node/.style = {minimum width=\MinWidth cm, minimum height=\MinWidth cm, inner sep=0,outer sep=0},
        ]

            \node[Node] at (\baseX + 0 * \MinWidth,\baseY + \MinWidth) {\scalebox{\MinWidth}{128}};
            \node[Node] at (\baseX + 1 * \MinWidth,\baseY + \MinWidth) {\scalebox{\MinWidth}{256}};
            \node[Node] at (\baseX + 2 * \MinWidth,\baseY + \MinWidth) {\scalebox{\MinWidth}{512}};
            \node[Node] at (\baseX + 3 * \MinWidth,\baseY + \MinWidth) {\scalebox{\MinWidth}{1024}};
            \node[Node] at (\baseX + 4 * \MinWidth,\baseY + \MinWidth) {\scalebox{\MinWidth}{2048}};
            \node[Node] at (\baseX + 5 * \MinWidth,\baseY + \MinWidth) {\scalebox{\MinWidth}{4096}};
            \node[Node] at (\baseX + 6 * \MinWidth,\baseY + \MinWidth) {\scalebox{\MinWidth}{8192}};
            
            \node[Node] at (\baseX - \MinWidth,\baseY - 0 * \MinWidth) {\scalebox{\MinWidth}{128}};
            \node[Node] at (\baseX - \MinWidth,\baseY - 1 * \MinWidth) {\scalebox{\MinWidth}{256}};
            \node[Node] at (\baseX - \MinWidth,\baseY - 2 * \MinWidth) {\scalebox{\MinWidth}{512}};
            \node[Node] at (\baseX - \MinWidth,\baseY - 3 * \MinWidth) {\scalebox{\MinWidth}{1024}};
            \node[Node] at (\baseX - \MinWidth,\baseY - 4 * \MinWidth) {\scalebox{\MinWidth}{2048}};
            \node[Node] at (\baseX - \MinWidth,\baseY - 5 * \MinWidth) {\scalebox{\MinWidth}{4096}};
            \node[Node] at (\baseX - \MinWidth,\baseY - 6 * \MinWidth) {\scalebox{\MinWidth}{8192}};

            \renewcommand{\baseY}{-1 * 0 * \MinWidth}%
            \CreateGradientColorCell{0.95}{(\baseX + 0 * \MinWidth,\baseY - 0.0)}{none}
            \CreateGradientColorCell{0.93}{(\baseX + 1 * \MinWidth,\baseY - 0.0)}{none}
            \CreateGradientColorCell{0.90}{(\baseX + 2 * \MinWidth,\baseY - 0.0)}{none}
            \CreateGradientColorCell{0.87}{(\baseX + 3 * \MinWidth,\baseY - 0.0)}{none}
            \CreateGradientColorCell{0.84}{(\baseX + 4 * \MinWidth,\baseY - 0.0)}{none}
            \CreateGradientColorCell{0.81}{(\baseX + 5 * \MinWidth,\baseY - 0.0)}{none}
            \CreateGradientColorCell{0.79}{(\baseX + 6 * \MinWidth,\baseY - 0.0)}{none}
            \renewcommand{\baseY}{-1 * 1 * \MinWidth}%
            \CreateGradientColorCell{0.94}{(\baseX + 0 * \MinWidth,\baseY - 0.0)}{none}
            \CreateGradientColorCell{0.90}{(\baseX + 1 * \MinWidth,\baseY - 0.0)}{none}
            \CreateGradientColorCell{0.86}{(\baseX + 2 * \MinWidth,\baseY - 0.0)}{none}
            \CreateGradientColorCell{0.84}{(\baseX + 3 * \MinWidth,\baseY - 0.0)}{none}
            \CreateGradientColorCell{0.81}{(\baseX + 4 * \MinWidth,\baseY - 0.0)}{none}
            \CreateGradientColorCell{0.79}{(\baseX + 5 * \MinWidth,\baseY - 0.0)}{none}
            \CreateGradientColorCell{0.78}{(\baseX + 6 * \MinWidth,\baseY - 0.0)}{none}
            \renewcommand{\baseY}{-1 * 2 * \MinWidth}%
            \CreateGradientColorCell{0.90}{(\baseX + 0 * \MinWidth,\baseY - 0.0)}{none}
            \CreateGradientColorCell{0.87}{(\baseX + 1 * \MinWidth,\baseY - 0.0)}{none}
            \CreateGradientColorCell{0.85}{(\baseX + 2 * \MinWidth,\baseY - 0.0)}{none}
            \CreateGradientColorCell{0.81}{(\baseX + 3 * \MinWidth,\baseY - 0.0)}{none}
            \CreateGradientColorCell{0.79}{(\baseX + 4 * \MinWidth,\baseY - 0.0)}{none}
            \CreateGradientColorCell{0.78}{(\baseX + 5 * \MinWidth,\baseY - 0.0)}{none}
            \CreateGradientColorCell{1.37}{(\baseX + 6 * \MinWidth,\baseY - 0.0)}{none}
            \renewcommand{\baseY}{-1 * 3 * \MinWidth}%
            \CreateGradientColorCell{0.87}{(\baseX + 0 * \MinWidth,\baseY - 0.0)}{none}
            \CreateGradientColorCell{0.84}{(\baseX + 1 * \MinWidth,\baseY - 0.0)}{none}
            \CreateGradientColorCell{0.82}{(\baseX + 2 * \MinWidth,\baseY - 0.0)}{none}
            \CreateGradientColorCell{0.80}{(\baseX + 3 * \MinWidth,\baseY - 0.0)}{none}
            \CreateGradientColorCell{0.78}{(\baseX + 4 * \MinWidth,\baseY - 0.0)}{none}
            \CreateGradientColorCell{1.31}{(\baseX + 5 * \MinWidth,\baseY - 0.0)}{none}
            \CreateGradientColorCell{2.71}{(\baseX + 6 * \MinWidth,\baseY - 0.0)}{none}
            \renewcommand{\baseY}{-1 * 4 * \MinWidth}%
            \CreateGradientColorCell{0.85}{(\baseX + 0 * \MinWidth,\baseY - 0.0)}{none}
            \CreateGradientColorCell{0.82}{(\baseX + 1 * \MinWidth,\baseY - 0.0)}{none}
            \CreateGradientColorCell{0.80}{(\baseX + 2 * \MinWidth,\baseY - 0.0)}{none}
            \CreateGradientColorCell{0.79}{(\baseX + 3 * \MinWidth,\baseY - 0.0)}{none}
            \CreateGradientColorCell{1.31}{(\baseX + 4 * \MinWidth,\baseY - 0.0)}{none}
            \CreateGradientColorCell{2.70}{(\baseX + 5 * \MinWidth,\baseY - 0.0)}{none}
            \CreateGradientColorCell{2.24}{(\baseX + 6 * \MinWidth,\baseY - 0.0)}{none}
            \renewcommand{\baseY}{-1 * 5 * \MinWidth}%
            \CreateGradientColorCell{0.82}{(\baseX + 0 * \MinWidth,\baseY - 0.0)}{none}
            \CreateGradientColorCell{0.80}{(\baseX + 1 * \MinWidth,\baseY - 0.0)}{none}
            \CreateGradientColorCell{0.79}{(\baseX + 2 * \MinWidth,\baseY - 0.0)}{none}
            \CreateGradientColorCell{1.35}{(\baseX + 3 * \MinWidth,\baseY - 0.0)}{none}
            \CreateGradientColorCell{2.71}{(\baseX + 4 * \MinWidth,\baseY - 0.0)}{none}
            \CreateGradientColorCell{2.22}{(\baseX + 5 * \MinWidth,\baseY - 0.0)}{none}
            \CreateGradientColorCell{1.53}{(\baseX + 6 * \MinWidth,\baseY - 0.0)}{none}
            \renewcommand{\baseY}{-1 * 6 * \MinWidth}%
            \CreateGradientColorCell{0.80}{(\baseX + 0 * \MinWidth,\baseY - 0.0)}{none}
            \CreateGradientColorCell{0.79}{(\baseX + 1 * \MinWidth,\baseY - 0.0)}{none}
            \CreateGradientColorCell{1.48}{(\baseX + 2 * \MinWidth,\baseY - 0.0)}{none}
            \CreateGradientColorCell{2.73}{(\baseX + 3 * \MinWidth,\baseY - 0.0)}{none}
            \CreateGradientColorCell{2.15}{(\baseX + 4 * \MinWidth,\baseY - 0.0)}{none}
            \CreateGradientColorCell{1.52}{(\baseX + 5 * \MinWidth,\baseY - 0.0)}{\HighlighColor}
            \CreateGradientColorCell{1.52}{(\baseX + 6 * \MinWidth,\baseY - 0.0)}{none}

            \draw[->] (-1 * \MinWidth / 2,\MinWidth / 2) -- (\MinWidth * 6.5,\MinWidth / 2);
            \node[minimum height=\MinWidth cm, inner sep=0,outer sep=0] at (3 * \MinWidth,1.5 * \MinWidth) {\scalebox{\MinWidth}{ Input Size}};

            \draw[->] (-1 * \MinWidth / 2,\MinWidth / 2) -- (-1 * \MinWidth / 2,-1 * \MinWidth * 6.5);
            \node[minimum height=\MinWidth cm, inner sep=0,outer sep=0] at (-0.4 * \MinWidth,-7 * \MinWidth) {\scalebox{\MinWidth}{ Output Size}};
        \end{tikzpicture}
        \vspace{-3mm}
        \caption{
            \centering
            Execution Time of \ourmethodWA{W1A1}
        }
        \label{fig:results:DifferentBitwidths:ourW1A1}
    \end{subfigure}
	\begin{subfigure}[b]{0.49\linewidth}
        \centering
        \newcommand*{\MaxNumber}{1.313}
        \newcommand*{\MidNumber}{1}
        \newcommand*{\MinNumber}{0.965}
        \newcommand*{\MaxColor}{red}
        \newcommand*{\MidColor}{white}
        \newcommand*{\MinColor}{maxGreen}
        \begin{tikzpicture}[
            Node/.style = {minimum width=\MinWidth cm, minimum height=\MinWidth cm, inner sep=0,outer sep=0},
        ]
            \node[Node] at (\baseX + 0 * \MinWidth,\baseY + \MinWidth) {\scalebox{\MinWidth}{128}};
            \node[Node] at (\baseX + 1 * \MinWidth,\baseY + \MinWidth) {\scalebox{\MinWidth}{256}};
            \node[Node] at (\baseX + 2 * \MinWidth,\baseY + \MinWidth) {\scalebox{\MinWidth}{512}};
            \node[Node] at (\baseX + 3 * \MinWidth,\baseY + \MinWidth) {\scalebox{\MinWidth}{1024}};
            \node[Node] at (\baseX + 4 * \MinWidth,\baseY + \MinWidth) {\scalebox{\MinWidth}{2048}};
            \node[Node] at (\baseX + 5 * \MinWidth,\baseY + \MinWidth) {\scalebox{\MinWidth}{4096}};
            \node[Node] at (\baseX + 6 * \MinWidth,\baseY + \MinWidth) {\scalebox{\MinWidth}{8192}};
            
            \node[Node] at (\baseX - \MinWidth,\baseY - 0 * \MinWidth) {\scalebox{\MinWidth}{128}};
            \node[Node] at (\baseX - \MinWidth,\baseY - 1 * \MinWidth) {\scalebox{\MinWidth}{256}};
            \node[Node] at (\baseX - \MinWidth,\baseY - 2 * \MinWidth) {\scalebox{\MinWidth}{512}};
            \node[Node] at (\baseX - \MinWidth,\baseY - 3 * \MinWidth) {\scalebox{\MinWidth}{1024}};
            \node[Node] at (\baseX - \MinWidth,\baseY - 4 * \MinWidth) {\scalebox{\MinWidth}{2048}};
            \node[Node] at (\baseX - \MinWidth,\baseY - 5 * \MinWidth) {\scalebox{\MinWidth}{4096}};
            \node[Node] at (\baseX - \MinWidth,\baseY - 6 * \MinWidth) {\scalebox{\MinWidth}{8192}};

            \renewcommand{\baseY}{-1 * 0 * \MinWidth}%
            \CreateGradientColorCell{0.988}{(\baseX + 0 * \MinWidth,\baseY - 0.0)}{none}
            \CreateGradientColorCell{0.984}{(\baseX + 1 * \MinWidth,\baseY - 0.0)}{none}
            \CreateGradientColorCell{0.980}{(\baseX + 2 * \MinWidth,\baseY - 0.0)}{none}
            \CreateGradientColorCell{0.973}{(\baseX + 3 * \MinWidth,\baseY - 0.0)}{none}
            \CreateGradientColorCell{0.970}{(\baseX + 4 * \MinWidth,\baseY - 0.0)}{none}
            \CreateGradientColorCell{0.967}{(\baseX + 5 * \MinWidth,\baseY - 0.0)}{none}
            \CreateGradientColorCell{0.966}{(\baseX + 6 * \MinWidth,\baseY - 0.0)}{none}
            \renewcommand{\baseY}{-1 * 1 * \MinWidth}%
            \CreateGradientColorCell{0.980}{(\baseX + 0 * \MinWidth,\baseY - 0.0)}{none}
            \CreateGradientColorCell{0.979}{(\baseX + 1 * \MinWidth,\baseY - 0.0)}{none}
            \CreateGradientColorCell{0.974}{(\baseX + 2 * \MinWidth,\baseY - 0.0)}{none}
            \CreateGradientColorCell{0.970}{(\baseX + 3 * \MinWidth,\baseY - 0.0)}{none}
            \CreateGradientColorCell{0.968}{(\baseX + 4 * \MinWidth,\baseY - 0.0)}{none}
            \CreateGradientColorCell{0.966}{(\baseX + 5 * \MinWidth,\baseY - 0.0)}{none}
            \CreateGradientColorCell{0.966}{(\baseX + 6 * \MinWidth,\baseY - 0.0)}{none}
            \renewcommand{\baseY}{-1 * 2 * \MinWidth}%
            \CreateGradientColorCell{0.980}{(\baseX + 0 * \MinWidth,\baseY - 0.0)}{none}
            \CreateGradientColorCell{0.974}{(\baseX + 1 * \MinWidth,\baseY - 0.0)}{none}
            \CreateGradientColorCell{0.971}{(\baseX + 2 * \MinWidth,\baseY - 0.0)}{none}
            \CreateGradientColorCell{0.968}{(\baseX + 3 * \MinWidth,\baseY - 0.0)}{none}
            \CreateGradientColorCell{0.966}{(\baseX + 4 * \MinWidth,\baseY - 0.0)}{none}
            \CreateGradientColorCell{0.966}{(\baseX + 5 * \MinWidth,\baseY - 0.0)}{none}
            \CreateGradientColorCell{0.965}{(\baseX + 6 * \MinWidth,\baseY - 0.0)}{none}
            \renewcommand{\baseY}{-1 * 3 * \MinWidth}%
            \CreateGradientColorCell{0.976}{(\baseX + 0 * \MinWidth,\baseY - 0.0)}{none}
            \CreateGradientColorCell{0.971}{(\baseX + 1 * \MinWidth,\baseY - 0.0)}{none}
            \CreateGradientColorCell{0.969}{(\baseX + 2 * \MinWidth,\baseY - 0.0)}{none}
            \CreateGradientColorCell{0.967}{(\baseX + 3 * \MinWidth,\baseY - 0.0)}{none}
            \CreateGradientColorCell{0.966}{(\baseX + 4 * \MinWidth,\baseY - 0.0)}{none}
            \CreateGradientColorCell{0.966}{(\baseX + 5 * \MinWidth,\baseY - 0.0)}{none}
            \CreateGradientColorCell{0.965}{(\baseX + 6 * \MinWidth,\baseY - 0.0)}{none}
            \renewcommand{\baseY}{-1 * 4 * \MinWidth}%
            \CreateGradientColorCell{0.973}{(\baseX + 0 * \MinWidth,\baseY - 0.0)}{none}
            \CreateGradientColorCell{0.969}{(\baseX + 1 * \MinWidth,\baseY - 0.0)}{none}
            \CreateGradientColorCell{0.967}{(\baseX + 2 * \MinWidth,\baseY - 0.0)}{none}
            \CreateGradientColorCell{0.966}{(\baseX + 3 * \MinWidth,\baseY - 0.0)}{none}
            \CreateGradientColorCell{0.966}{(\baseX + 4 * \MinWidth,\baseY - 0.0)}{none}
            \CreateGradientColorCell{0.965}{(\baseX + 5 * \MinWidth,\baseY - 0.0)}{none}
            \CreateGradientColorCell{0.965}{(\baseX + 6 * \MinWidth,\baseY - 0.0)}{none}
            \renewcommand{\baseY}{-1 * 5 * \MinWidth}%
            \CreateGradientColorCell{0.971}{(\baseX + 0 * \MinWidth,\baseY - 0.0)}{none}
            \CreateGradientColorCell{0.968}{(\baseX + 1 * \MinWidth,\baseY - 0.0)}{none}
            \CreateGradientColorCell{0.967}{(\baseX + 2 * \MinWidth,\baseY - 0.0)}{none}
            \CreateGradientColorCell{0.966}{(\baseX + 3 * \MinWidth,\baseY - 0.0)}{none}
            \CreateGradientColorCell{0.966}{(\baseX + 4 * \MinWidth,\baseY - 0.0)}{none}
            \CreateGradientColorCell{0.965}{(\baseX + 5 * \MinWidth,\baseY - 0.0)}{none}
            \CreateGradientColorCell{0.965}{(\baseX + 6 * \MinWidth,\baseY - 0.0)}{none}
            \renewcommand{\baseY}{-1 * 6 * \MinWidth}%
            \CreateGradientColorCell{0.970}{(\baseX + 0 * \MinWidth,\baseY - 0.0)}{none}
            \CreateGradientColorCell{0.968}{(\baseX + 1 * \MinWidth,\baseY - 0.0)}{none}
            \CreateGradientColorCell{0.966}{(\baseX + 2 * \MinWidth,\baseY - 0.0)}{none}
            \CreateGradientColorCell{0.966}{(\baseX + 3 * \MinWidth,\baseY - 0.0)}{none}
            \CreateGradientColorCell{0.965}{(\baseX + 4 * \MinWidth,\baseY - 0.0)}{none}
            \CreateGradientColorCell{0.965}{(\baseX + 5 * \MinWidth,\baseY - 0.0)}{\HighlighColor}
            \CreateGradientColorCell{0.965}{(\baseX + 6 * \MinWidth,\baseY - 0.0)}{none}

            \draw[->] (-1 * \MinWidth / 2,\MinWidth / 2) -- (\MinWidth * 6.5,\MinWidth / 2);
            \node[minimum height=\MinWidth cm, inner sep=0,outer sep=0] at (3 * \MinWidth,1.5 * \MinWidth) {\scalebox{\MinWidth}{ Input Size}};

            \draw[->] (-1 * \MinWidth / 2,\MinWidth / 2) -- (-1 * \MinWidth / 2,-1 * \MinWidth * 6.5);
            \node[minimum height=\MinWidth cm, inner sep=0,outer sep=0] at (-0.4 * \MinWidth,-7 * \MinWidth) {\scalebox{\MinWidth}{ Output Size}};
        \end{tikzpicture}
        \vspace{-3mm}
        \caption{
            \centering
            Instructions Count of \ourmethodWA{W2A2}
        }
        \label{fig:results:DifferentBitwidths:inst:ourW2A2}
    \end{subfigure}
    \begin{subfigure}[b]{0.49\linewidth}
        \centering
        \newcommand*{\MaxNumber}{1.313}
        \newcommand*{\MidNumber}{1}
        \newcommand*{\MinNumber}{0.965}
        \newcommand*{\MaxColor}{red}
        \newcommand*{\MidColor}{white}
        \newcommand*{\MinColor}{maxGreen}
        \begin{tikzpicture}[
            Node/.style = {minimum width=\MinWidth cm, minimum height=\MinWidth cm, inner sep=0,outer sep=0},
        ]
            \node[Node] at (\baseX + 0 * \MinWidth,\baseY + \MinWidth) {\scalebox{\MinWidth}{128}};
            \node[Node] at (\baseX + 1 * \MinWidth,\baseY + \MinWidth) {\scalebox{\MinWidth}{256}};
            \node[Node] at (\baseX + 2 * \MinWidth,\baseY + \MinWidth) {\scalebox{\MinWidth}{512}};
            \node[Node] at (\baseX + 3 * \MinWidth,\baseY + \MinWidth) {\scalebox{\MinWidth}{1024}};
            \node[Node] at (\baseX + 4 * \MinWidth,\baseY + \MinWidth) {\scalebox{\MinWidth}{2048}};
            \node[Node] at (\baseX + 5 * \MinWidth,\baseY + \MinWidth) {\scalebox{\MinWidth}{4096}};
            \node[Node] at (\baseX + 6 * \MinWidth,\baseY + \MinWidth) {\scalebox{\MinWidth}{8192}};
            
            \node[Node] at (\baseX - \MinWidth,\baseY - 0 * \MinWidth) {\scalebox{\MinWidth}{128}};
            \node[Node] at (\baseX - \MinWidth,\baseY - 1 * \MinWidth) {\scalebox{\MinWidth}{256}};
            \node[Node] at (\baseX - \MinWidth,\baseY - 2 * \MinWidth) {\scalebox{\MinWidth}{512}};
            \node[Node] at (\baseX - \MinWidth,\baseY - 3 * \MinWidth) {\scalebox{\MinWidth}{1024}};
            \node[Node] at (\baseX - \MinWidth,\baseY - 4 * \MinWidth) {\scalebox{\MinWidth}{2048}};
            \node[Node] at (\baseX - \MinWidth,\baseY - 5 * \MinWidth) {\scalebox{\MinWidth}{4096}};
            \node[Node] at (\baseX - \MinWidth,\baseY - 6 * \MinWidth) {\scalebox{\MinWidth}{8192}};

            \renewcommand{\baseY}{-1 * 0 * \MinWidth}%
            \CreateGradientColorCell{1.082}{(\baseX + 0 * \MinWidth,\baseY - 0.0)}{none}
            \CreateGradientColorCell{1.126}{(\baseX + 1 * \MinWidth,\baseY - 0.0)}{none}
            \CreateGradientColorCell{1.180}{(\baseX + 2 * \MinWidth,\baseY - 0.0)}{none}
            \CreateGradientColorCell{1.227}{(\baseX + 3 * \MinWidth,\baseY - 0.0)}{none}
            \CreateGradientColorCell{1.262}{(\baseX + 4 * \MinWidth,\baseY - 0.0)}{none}
            \CreateGradientColorCell{1.284}{(\baseX + 5 * \MinWidth,\baseY - 0.0)}{none}
            \CreateGradientColorCell{1.297}{(\baseX + 6 * \MinWidth,\baseY - 0.0)}{none}
            \renewcommand{\baseY}{-1 * 1 * \MinWidth}%
            \CreateGradientColorCell{1.124}{(\baseX + 0 * \MinWidth,\baseY - 0.0)}{none}
            \CreateGradientColorCell{1.174}{(\baseX + 1 * \MinWidth,\baseY - 0.0)}{none}
            \CreateGradientColorCell{1.224}{(\baseX + 2 * \MinWidth,\baseY - 0.0)}{none}
            \CreateGradientColorCell{1.262}{(\baseX + 3 * \MinWidth,\baseY - 0.0)}{none}
            \CreateGradientColorCell{1.284}{(\baseX + 4 * \MinWidth,\baseY - 0.0)}{none}
            \CreateGradientColorCell{1.298}{(\baseX + 5 * \MinWidth,\baseY - 0.0)}{none}
            \CreateGradientColorCell{1.305}{(\baseX + 6 * \MinWidth,\baseY - 0.0)}{none}
            \renewcommand{\baseY}{-1 * 2 * \MinWidth}%
            \CreateGradientColorCell{1.167}{(\baseX + 0 * \MinWidth,\baseY - 0.0)}{none}
            \CreateGradientColorCell{1.217}{(\baseX + 1 * \MinWidth,\baseY - 0.0)}{none}
            \CreateGradientColorCell{1.259}{(\baseX + 2 * \MinWidth,\baseY - 0.0)}{none}
            \CreateGradientColorCell{1.283}{(\baseX + 3 * \MinWidth,\baseY - 0.0)}{none}
            \CreateGradientColorCell{1.297}{(\baseX + 4 * \MinWidth,\baseY - 0.0)}{none}
            \CreateGradientColorCell{1.305}{(\baseX + 5 * \MinWidth,\baseY - 0.0)}{none}
            \CreateGradientColorCell{1.309}{(\baseX + 6 * \MinWidth,\baseY - 0.0)}{none}
            \renewcommand{\baseY}{-1 * 3 * \MinWidth}%
            \CreateGradientColorCell{1.212}{(\baseX + 0 * \MinWidth,\baseY - 0.0)}{none}
            \CreateGradientColorCell{1.253}{(\baseX + 1 * \MinWidth,\baseY - 0.0)}{none}
            \CreateGradientColorCell{1.279}{(\baseX + 2 * \MinWidth,\baseY - 0.0)}{none}
            \CreateGradientColorCell{1.296}{(\baseX + 3 * \MinWidth,\baseY - 0.0)}{none}
            \CreateGradientColorCell{1.304}{(\baseX + 4 * \MinWidth,\baseY - 0.0)}{none}
            \CreateGradientColorCell{1.309}{(\baseX + 5 * \MinWidth,\baseY - 0.0)}{none}
            \CreateGradientColorCell{1.311}{(\baseX + 6 * \MinWidth,\baseY - 0.0)}{none}
            \renewcommand{\baseY}{-1 * 4 * \MinWidth}%
            \CreateGradientColorCell{1.243}{(\baseX + 0 * \MinWidth,\baseY - 0.0)}{none}
            \CreateGradientColorCell{1.274}{(\baseX + 1 * \MinWidth,\baseY - 0.0)}{none}
            \CreateGradientColorCell{1.292}{(\baseX + 2 * \MinWidth,\baseY - 0.0)}{none}
            \CreateGradientColorCell{1.303}{(\baseX + 3 * \MinWidth,\baseY - 0.0)}{none}
            \CreateGradientColorCell{1.308}{(\baseX + 4 * \MinWidth,\baseY - 0.0)}{none}
            \CreateGradientColorCell{1.311}{(\baseX + 5 * \MinWidth,\baseY - 0.0)}{none}
            \CreateGradientColorCell{1.312}{(\baseX + 6 * \MinWidth,\baseY - 0.0)}{none}
            \renewcommand{\baseY}{-1 * 5 * \MinWidth}%
            \CreateGradientColorCell{1.261}{(\baseX + 0 * \MinWidth,\baseY - 0.0)}{none}
            \CreateGradientColorCell{1.285}{(\baseX + 1 * \MinWidth,\baseY - 0.0)}{none}
            \CreateGradientColorCell{1.299}{(\baseX + 2 * \MinWidth,\baseY - 0.0)}{none}
            \CreateGradientColorCell{1.306}{(\baseX + 3 * \MinWidth,\baseY - 0.0)}{none}
            \CreateGradientColorCell{1.310}{(\baseX + 4 * \MinWidth,\baseY - 0.0)}{none}
            \CreateGradientColorCell{1.312}{(\baseX + 5 * \MinWidth,\baseY - 0.0)}{none}
            \CreateGradientColorCell{1.313}{(\baseX + 6 * \MinWidth,\baseY - 0.0)}{none}
            \renewcommand{\baseY}{-1 * 6 * \MinWidth}%
            \CreateGradientColorCell{1.272}{(\baseX + 0 * \MinWidth,\baseY - 0.0)}{none}
            \CreateGradientColorCell{1.291}{(\baseX + 1 * \MinWidth,\baseY - 0.0)}{none}
            \CreateGradientColorCell{1.302}{(\baseX + 2 * \MinWidth,\baseY - 0.0)}{none}
            \CreateGradientColorCell{1.308}{(\baseX + 3 * \MinWidth,\baseY - 0.0)}{none}
            \CreateGradientColorCell{1.311}{(\baseX + 4 * \MinWidth,\baseY - 0.0)}{none}
            \CreateGradientColorCell{1.312}{(\baseX + 5 * \MinWidth,\baseY - 0.0)}{\HighlighColor}
            \CreateGradientColorCell{1.313}{(\baseX + 6 * \MinWidth,\baseY - 0.0)}{none}

            \draw[->] (-1 * \MinWidth / 2,\MinWidth / 2) -- (\MinWidth * 6.5,\MinWidth / 2);
            \node[minimum height=\MinWidth cm, inner sep=0,outer sep=0] at (3 * \MinWidth,1.5 * \MinWidth) {\scalebox{\MinWidth}{ Input Size}};

            \draw[->] (-1 * \MinWidth / 2,\MinWidth / 2) -- (-1 * \MinWidth / 2,-1 * \MinWidth * 6.5);
            \node[minimum height=\MinWidth cm, inner sep=0,outer sep=0] at (-0.4 * \MinWidth,-7 * \MinWidth) {\scalebox{\MinWidth}{ Output Size}};
        \end{tikzpicture}
        \vspace{-3mm}
        \caption{
            \centering
            Instructions Count of \ourmethodWA{W1A1}
        }
        \label{fig:results:DifferentBitwidths:inst:ourW1A1}
    \end{subfigure}
    \centering
    \vspace{-3mm}
    \caption{
        \centering
        The effect of different quantization bit-widths on the speedups aginast $W4A4$ ($T_{W4A4}$ / $T_{case}$)
		and the increase in executed Instructions Count w.r.t. $W4A4$ ($I_{case}$ / $I_{W4A4}$) obtained by our
		method.
    }
    \vspace{-4mm}
    \label{fig:results:DifferentBitwidths:mainFig}
\end{figure}
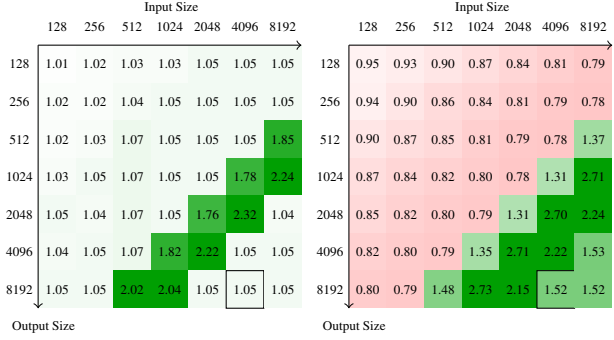
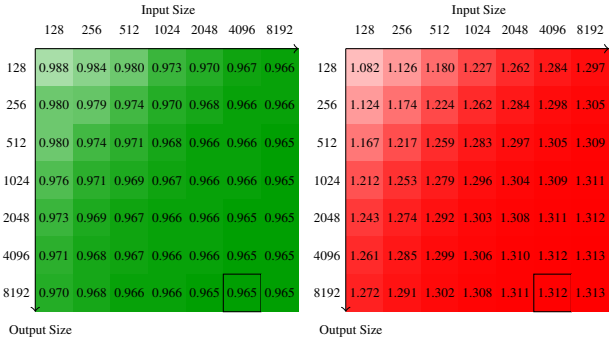

%% file: figures/mozilla-deepspeech-arch.tex
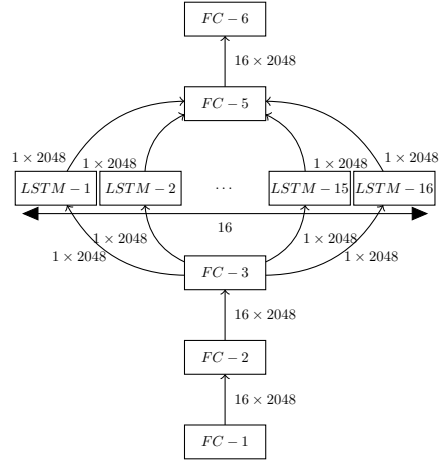
\begin{figure}[t]
    \centering
    \newcommand*{\HighlighColor}{black}%
    \newcommand{\baseX}{0}%
    \newcommand{\baseY}{0}%
    \newcommand*{\MinWidth}{0.45}%
    \newcommand*{\nodeScale}{0.55}%
    \newcommand{\scaleNode}[3]{\scalebox{\nodeScale}{#1}}%
    \begin{tikzpicture}[
        Node/.style = {
            draw, 
            minimum width=2.4 * \MinWidth cm, 
            minimum height=\MinWidth cm, 
            inner sep=0,outer sep=0,
            align=center,
        },
    ]
        \node[Node] (FC1N) at (0,0.0 * 2.5 * \MinWidth) {\scalebox{\nodeScale}{$FC-1$}};
        \node[Node] (FC2N) at (0,1.0 * 2.5 * \MinWidth) {\scalebox{\nodeScale}{$FC-2$}};
        \node[Node] (FC3N) at (0,2.0 * 2.5 * \MinWidth) {\scalebox{\nodeScale}{$FC-3$}};

        \node[Node] (LSTM1N)  at (-2 * 2.5 * \MinWidth ,3.0 * 2.5 * \MinWidth) {\scalebox{\nodeScale}{$LSTM-1$}};
        \node[Node] (LSTM2N)  at (-1 * 2.5 * \MinWidth ,3.0 * 2.5 * \MinWidth) {\scalebox{\nodeScale}{$LSTM-2$}};
        \node[Node, draw=none] (LSTMDN)  at ( 0 * 2.5 * \MinWidth ,3.0 * 2.5 * \MinWidth) {\scalebox{\nodeScale}{$\dots$}};
        \node[Node] (LSTM15N) at ( 1 * 2.5 * \MinWidth ,3.0 * 2.5 * \MinWidth) {\scalebox{\nodeScale}{$LSTM-15$}};
        \node[Node] (LSTM16N) at ( 2 * 2.5 * \MinWidth ,3.0 * 2.5 * \MinWidth) {\scalebox{\nodeScale}{$LSTM-16$}};

        \node[Node] (FC5N) at (0,4.0 * 2.5 * \MinWidth) {\scalebox{\nodeScale}{$FC-5$}};
        \node[Node] (FC6N) at (0,5.0 * 2.5 * \MinWidth) {\scalebox{\nodeScale}{$FC-6$}};

        \draw[->] (FC1N) -- (FC2N) node[midway,right,rotate=0] {\scalebox{\nodeScale}{$16\times2048$}};
        \draw[->] (FC2N) -- (FC3N) node[midway,right,rotate=0] {\scalebox{\nodeScale}{$16\times2048$}};

        \draw[->]
            (FC3N) edge[bend left ] node [left,  midway] {\scalebox{\nodeScale}{$1\times2048$}} (LSTM1N)
            (FC3N) edge[bend left ] node [left,  midway] {\scalebox{\nodeScale}{$1\times2048$}} (LSTM2N)
            (FC3N) edge[bend right] node [right, midway] {\scalebox{\nodeScale}{$1\times2048$}} (LSTM15N)
            (FC3N) edge[bend right] node [right, midway] {\scalebox{\nodeScale}{$1\times2048$}} (LSTM16N)
        ;

        \draw[->]
            (LSTM1N)  edge[bend left ] node [left,  pos=0.1] {\scalebox{\nodeScale}{$1\times2048$}} (FC5N)
            (LSTM2N)  edge[bend left ] node [left,  pos=0.1] {\scalebox{\nodeScale}{$1\times2048$}} (FC5N)
            (LSTM15N) edge[bend right] node [right, pos=0.1] {\scalebox{\nodeScale}{$1\times2048$}} (FC5N)
            (LSTM16N) edge[bend right] node [right, pos=0.1] {\scalebox{\nodeScale}{$1\times2048$}} (FC5N)
        ;

        \draw[->] (FC5N) -- (FC6N) node[midway,right,rotate=0] {\scalebox{\nodeScale}{$16\times2048$}};

        \draw[>=triangle 45, <->] (-2 * 2.5 * \MinWidth - \MinWidth, 3.0 * 2.5 * \MinWidth - 0.75 * \MinWidth) -- 
                                    ( 2 * 2.5 * \MinWidth + \MinWidth, 3.0 * 2.5 * \MinWidth - 0.75 * \MinWidth) 
        node[midway,below] {\scalebox{\nodeScale}{$16$}};

    \end{tikzpicture}
    \centering
    \vspace{-3mm}
    \caption{
        The network architecture of Mozilla DeepSpeech.
    }
    \vspace{-3mm}
    \label{fig:results:endToEnd:MozillaDeepSpeech}
\end{figure}

%% file: figures/end-to-end-results.tex
\begin{figure*}[t!]
    \centering
        \begin{footnotesize}
            \begin{tikzpicture}
                \begin{axis}[
                    ybar=0pt,
                    ymin=-1,
                    ymax=4.5,
                    width=0.995\linewidth,
                    height=0.625*\axisdefaultheight,
                    symbolic x coords={
                        Layer 1, %
                        Layer 2, %
                        Layer 3, %
                        LSTM, %
                        Layer 5, %
                        Layer 6, %
                        Total
                    },
                    tick label style={font=\footnotesize},
                    label style={font=\footnotesize},
                    ytick={-1,-0.5,0,0.5,1.5,1,2.5,2,3.5,3,4,4.5},
                    xtick = {
                        Layer 1, %
                        Layer 2, %
                        Layer 3, %
                        LSTM, %
                        Layer 5, %
                        Layer 6, %
                        Total
                    },
                    ylabel={Log$_{10}$ Execution Time (millisecond)},
                    nodes near coords align={vertical},
                    legend pos=north east,
                    bar width=0.1cm,
                    align={center},
                    ymajorgrids,
                    xminorgrids = true,
                    minor tick num=1,
                    legend columns=5,
                    legend style={at={(0.49,-0.2)},anchor=north,font=\footnotesize}
                    ]
                    \addplot[C1, fill=C1, postaction={pattern=\COnePattern}]
                    coordinates {(Layer 1,0.464) (Layer 2,1.692) (Layer 3,1.691)
                                (LSTM,3.502) (Layer 5,1.691) (Layer 6,-0.660)
                                (Total,3.526)};
                    
                    \addplot[C2, fill=C2, postaction={pattern=\CTowPattern}] 
                    coordinates {(Layer 1,0.464) (Layer 2,1.692) (Layer 3,1.691)
                                (LSTM,3.502) (Layer 5,1.691) (Layer 6,-0.660)
                                (Total,3.526)};
                    
                    \addplot[C3, fill=C3, postaction={pattern=\CThreePattern}] 
                    coordinates {(Layer 1,0.464) (Layer 2,1.760) (Layer 3,1.759)
                                (LSTM,3.569) (Layer 5,1.759) (Layer 6,-0.662)
                                (Total,3.592)};

                    \addplot[C4, fill=C4, postaction={pattern=\CFourPattern}] 
                    coordinates {(Layer 1,0.475) (Layer 2,1.043) (Layer 3,1.040)
                                 (LSTM,2.134) (Layer 5,1.046) (Layer 6,-0.664)
                                 (Total,2.301)};
                    
                    \addplot[C5, fill=C5, postaction={pattern=\CFivePattern}] 
                    coordinates {(Layer 1,0.475) (Layer 2,1.043) (Layer 3,1.040)
                                 (LSTM,2.294) (Layer 5,1.046) (Layer 6,-0.666)
                                 (Total,2.416)};
                    
                    \addplot[C6, fill=C6, postaction={pattern=\CSixPattern}] 
                    coordinates {(Layer 1,0.474) (Layer 2,1.043) (Layer 3,1.041)
                                 (LSTM,2.315) (Layer 5,1.046) (Layer 6,-0.660)
                                 (Total,2.433)};
                    
                    \addplot[C7, fill=C7, postaction={pattern=\CSevenPattern}] 
                    coordinates {(Layer 1,0.474) (Layer 2,1.043) (Layer 3,1.040)
                                (LSTM,2.555) (Layer 5,1.045) (Layer 6,-0.666)
                                (Total,2.626)};
                    
                    \addplot[C8, fill=C8, postaction={pattern=\CEightPattern}] 
                    coordinates {(Total,2.524)};
                    
                    \addplot[C9, fill=C9, postaction={pattern=\CNinePattern}] 
                    coordinates {(Layer 1,0.475) (Layer 2,1.043) (Layer 3,1.040)
                                (LSTM,4.120) (Layer 5,1.046) (Layer 6,-0.670)
                                (Total,4.122)};
                    
                    \addplot[C10, fill=C10, postaction={pattern=\CTenPattern}] 
                    coordinates {(Layer 1,1.102) (Layer 2,1.621) (Layer 3,1.621)
                                (LSTM,3.643) (Layer 5,1.621) (Layer 6,-0.032)
                                (Total,3.659)};
                    
                    \addplot[C11, fill=C11, postaction={pattern=\CElevenPattern}] 
                    coordinates {(Layer 1,0.806) (Layer 2,1.433) (Layer 3,1.427)
                                (LSTM,3.179) (Layer 5,1.426) (Layer 6,-0.248)
                                (Total,3.209)};
                    
                    \addplot[C12, fill=C12, postaction={pattern=\CTwelvePattern}] 
                    coordinates {(Total,3.606)};
                    
                    \addplot[C13, fill=C13, postaction={pattern=\CThirteenPattern}] 
                    coordinates {(Layer 1,0.806) (Layer 2,1.433) (Layer 3,1.427)
                                (LSTM,3.487) (Layer 5,1.424) (Layer 6,-0.244)
                                (Total,3.501)};
                    
                    \addplot[C14, fill=C14, postaction={pattern=\CFourteenPattern}] 
                    coordinates {(Layer 1,2.017) (Layer 2,2.627) (Layer 3,2.628)
                                (LSTM,3.871) (Layer 5,2.627) (Layer 6,0.816)
                                (Total,3.946)};

                    \legend{
                        \ullpackwa{W1A1}    , \ullpackwa{W2A2}      , \ullpackwa{W3A3}  ,
                        \ourmethodwa{1}     , \ourmethodwa{2}       , \ourmethodwa{4}   ,
                        \ruyint{}           , \xnnpackint{}         , \tfliteint{}      , \gemmlowpint{},
                        \ruyfp{}            , \xnnpackfp{}          , \tflitefp{}       , \eigenfp{}
                    }
                \end{axis}
            \end{tikzpicture}
        \end{footnotesize}
        \vspace{-3mm}
    \caption{
        End-To-End evaluation on Mozilla DeepSpeech \cite{website:mozilladeepspeech} by per layer execution time breakdown, 
        for all of the methods except \xnnpack{}, because it does not allow per layer breakdown. \ourmethod{} does not support
        GEMM, so we used \ruyint{} for processing the GEMM operations.
    }
    \vspace{-3mm}
    \label{fig:results:endToEnd:deepspeechLayerBreakdown}
\end{figure*}
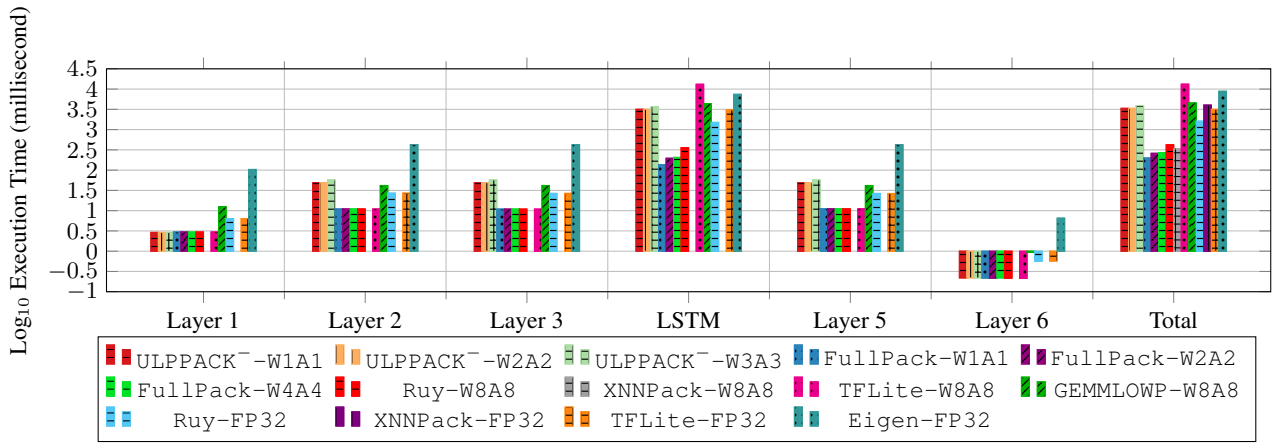

%% file: Sections/conclusion.tex
\section{Conclusion}
\label{conclusion}
    To tackle the bandwidth and capacity wastage of the latest solutions for sub-byte DNN models on constrained devices, we introduced a
    storage-processing co-design packing scheme for fixed-width vector instructions of commodity processors such as ARM's NEON architecture; 
    our solution needs no hardware extension, and fully utilizes the consumed memory bandwidth
    and memory footprint to respectively transfer and store only useful data.
    These packing schemes alongside their corresponding assembly kernels reduce expensive cache-misses, and thus improve performance despite needing 
    some additional vector instructions for unpacking the data in the vector registers.
    We implemented our scheme for the GEMV operation, common in fully-connected and other layers, of DNN models and provide it open source to the 
    community.
    We evaluated \ourmethod{} against nine other well-known techniques including the current state-of-the-art in the literature (ULPPACK)
    as well as industry (\ruy{}, \xnnpack{}, and \gemmlowp{}), and showed on cycle-accurate processor simulator that on average, \ourmethod{} consistently
    outperforms all rivals. \ourmethod{} achieves $2.35\times$ speedup against the baseline, \ruyint{}.
    For end-to-end evaluation, we applied all methods on Mozilla DeepSpeech and showed that \ourmethod{} outperforms all
    the others and provides $1.2-1.4\times$ speedup over the closest rival, \xnnpack{}.

%% file: Sections/appendix.tex
\section{On-Device Measurements}
\label{appendix:real-device-results}

    \input{figures/end-to-end-results-cnns-fcs-real-devices.tex}

    A complete demonstration of the results of our evaluation on fully connected layers of eleven different well-known CNN models are available in Figure~\ref{fig:appendix:endToEnd:CNNsFCs-rapsberry-pi-4}.

\section{Detailed Execution Metric Analysis}
\label{appendix:detailed-metrics}

    Here, we provide more detailed information about the execution of each method.
    As depicted in Figure~\ref{fig:results:performance:mainFig}, \ourmethod{} outperforms the baseline, \ruyint{}.
    One may expect that with narrower bit-width, the speedup of \ourmethod{} to get better, but as discussed in \S~\ref{results:WeightsVsActivaitons} and \S~\ref{results:DifferentBitwidths}, this is true only for \ourmethodwa{2} and not for \ourmethodwa{1}.
    As previously mentioned, the reason is that the instructions overhead becomes bottleneck in excution of \ourmethodwa{1}.
    For deeper inspection, we reported Instructions Count of each method for each model against the main baseline, \ruyint{} (Figure~\ref{fig:appendix:instructions:mainFig}).

    As we see, only when we quantize the activaitons, we still have $0.73\times$ instructions compared to the baseline.
    The reason is that \ruyint{} needs more preprocessing to prepare the data for processing with respect to \ourmethod{}.
    However, \xnnpackint{} needs way less instructions compared to both \ourmethod{} and \ruyint{}, $0.68\times$ of \ruyint{}.

    This, however, does not explain the reason behind why \ourmethod{} is faster than \xnnpackint{} for models with larger sizes.
    To investigate more, we evaluated Instructions Per Cycle (IPC) for each method.
    Figure~\ref{fig:appendix:ipc:mainFig} reports our results.
    Here, \ourmethod{} has better IPC than the baseline for almost all models and all sizes.
    However, if we compare the IPC of \xnnpackint{} with \ourmethod{} we can observe that \ourmethod{} has better IPC than the \xnnpackint{} for larger sizes of all the models, except $W8A4$,
    which causes \ourmethod{} to be faster than \xnnpackint{}, even with more executed instructions.  

    \input{figures/all-methods-instructions.tex}

    \input{figures/all-methods-ipc.tex}

%% file: figures/end-to-end-results-cnns-fcs-real-devices.tex
\begin{figure*}[t!]

    \begin{footnotesize}
        \begin{tikzpicture}
            \begin{axis}[
                ybar=0pt,
                ymin=0,
                ymax=2,
                width=0.995\linewidth,
                height=0.625*\axisdefaultheight,
                symbolic x coords={
                    DenseNet201,
                    EfficientNetV2L,
                    InceptionV3,
                    InceptionResNetV2,
                    MobileNetV2,
                    NASNetLarge,
                    RegNetY320,
                    ResNet152,
                    ResNet152V2,
                    VGG19,
                    Xception
                },
                tick label style={font=\footnotesize},
                label style={font=\footnotesize},
                ytick={0,0.5,1,1.5,2},
                xtick={
                    DenseNet201,
                    EfficientNetV2L,
                    InceptionV3,
                    InceptionResNetV2,
                    MobileNetV2,
                    NASNetLarge,
                    RegNetY320,
                    ResNet152,
                    ResNet152V2,
                    VGG19,
                    Xception
                },
                ylabel={Speedup},
                nodes near coords align={vertical},
                legend pos=north east,
                bar width=0.05cm,
                align={center},
                ymajorgrids,
                xminorgrids = true,
                minor tick num=1,
                legend columns=5,
                x tick label style={rotate=40,anchor=east},
                legend style={at={(0.49,-0.7)},anchor=north,font=\footnotesize}
                ]
                \addplot[C1, fill=C1, postaction={pattern=\COnePattern}]
                coordinates {
                    (DenseNet201,       0.02) (EfficientNetV2L,   0.02)
                    (InceptionV3,       0.02) (InceptionResNetV2, 0.02)
                    (MobileNetV2,       0.02) (NASNetLarge,       0.02)
                    (RegNetY320,        0.02) (ResNet152,         0.02)
                    (ResNet152V2,       0.02) (VGG19,             0.02)
                    (Xception,          0.02)
                };
                
                \addplot[C2, fill=C2, postaction={pattern=\CTowPattern}] 
                coordinates {
                    (DenseNet201,       0.02) (EfficientNetV2L,   0.02)
                    (InceptionV3,       0.02) (InceptionResNetV2, 0.02)
                    (MobileNetV2,       0.02) (NASNetLarge,       0.02)
                    (RegNetY320,        0.02) (ResNet152,         0.02)
                    (ResNet152V2,       0.02) (VGG19,             0.02)
                    (Xception,          0.02)
                };
                
                \addplot[C3, fill=C3, postaction={pattern=\CThreePattern}] 
                coordinates {
                    (DenseNet201,       0.02) (EfficientNetV2L,   0.02)
                    (InceptionV3,       0.02) (InceptionResNetV2, 0.02)
                    (MobileNetV2,       0.02) (NASNetLarge,       0.02)
                    (RegNetY320,        0.02) (ResNet152,         0.02)
                    (ResNet152V2,       0.02) (VGG19,             0.02)
                    (Xception,          0.02)
                };
                
                \addplot[C4, fill=C4, postaction={pattern=\CFourPattern}] 
                coordinates {
                    (DenseNet201,       1.11) (EfficientNetV2L,   1.21)
                    (InceptionV3,       1.38) (InceptionResNetV2, 1.18)
                    (MobileNetV2,       1.30) (NASNetLarge,       1.12)
                    (RegNetY320,        1.32) (ResNet152,         1.18)
                    (ResNet152V2,       1.20) (VGG19,             1.05)
                    (Xception,          1.20)
                };
                
                \addplot[C5, fill=C5, postaction={pattern=\CFivePattern}] 
                coordinates {
                    (DenseNet201,       1.37) (EfficientNetV2L,   1.48)
                    (InceptionV3,       1.69) (InceptionResNetV2, 1.47)
                    (MobileNetV2,       1.59) (NASNetLarge,       1.42)
                    (RegNetY320,        1.67) (ResNet152,         1.47)
                    (ResNet152V2,       1.48) (VGG19,             1.33)
                    (Xception,          1.48)
                };
                
                \addplot[C6, fill=C6, postaction={pattern=\CSixPattern}] 
                coordinates {
                    (DenseNet201,       1.33) (EfficientNetV2L,   1.39)
                    (InceptionV3,       1.62) (InceptionResNetV2, 1.40)
                    (MobileNetV2,       1.53) (NASNetLarge,       1.37)
                    (RegNetY320,        1.60) (ResNet152,         1.40)
                    (ResNet152V2,       1.40) (VGG19,             1.28)
                    (Xception,          1.43)
                };
                
                \addplot[C8, fill=C8, postaction={pattern=\CEightPattern}] 
                coordinates {
                    (DenseNet201,       1.24) (EfficientNetV2L,   1.37)
                    (InceptionV3,       1.59) (InceptionResNetV2, 1.27)
                    (MobileNetV2,       1.39) (NASNetLarge,       0.91)
                    (RegNetY320,        1.39) (ResNet152,         1.32)
                    (ResNet152V2,       1.30) (VGG19,             1.04)
                    (Xception,          1.35)
                };
                
                \addplot[C9, fill=C9, postaction={pattern=\CNinePattern}] 
                coordinates {
                    (DenseNet201,       0.01) (EfficientNetV2L,   0.01)
                    (InceptionV3,       0.01) (InceptionResNetV2, 0.01)
                    (MobileNetV2,       0.01) (NASNetLarge,       0.01)
                    (RegNetY320,        0.01) (ResNet152,         0.01)
                    (ResNet152V2,       0.01) (VGG19,             0.01)
                    (Xception,          0.01)
                };
                
                \addplot[C10, fill=C10, postaction={pattern=\CTenPattern}] 
                coordinates {
                    (DenseNet201,       0.80) (EfficientNetV2L,   0.86)
                    (InceptionV3,       1.17) (InceptionResNetV2, 0.86)
                    (MobileNetV2,       0.95) (NASNetLarge,       0.84)
                    (RegNetY320,        1.00) (ResNet152,         0.85)
                    (ResNet152V2,       0.84) (VGG19,             1.01)
                    (Xception,          0.86)
                };
                
                \addplot[C11, fill=C11, postaction={pattern=\CElevenPattern}] 
                coordinates {
                    (DenseNet201,       0.18) (EfficientNetV2L,   0.21)
                    (InceptionV3,       0.23) (InceptionResNetV2, 0.19)
                    (MobileNetV2,       0.22) (NASNetLarge,       0.17)
                    (RegNetY320,        0.21) (ResNet152,         0.19)
                    (ResNet152V2,       0.20) (VGG19,             0.02)
                    (Xception,          0.20)
                };
                
                \addplot[C12, fill=C12, postaction={pattern=\CTwelvePattern}] 
                coordinates {
                    (DenseNet201,       0.30) (EfficientNetV2L,   0.34)
                    (InceptionV3,       0.38) (InceptionResNetV2, 0.32)
                    (MobileNetV2,       0.37) (NASNetLarge,       0.29)
                    (RegNetY320,        0.34) (ResNet152,         0.32)
                    (ResNet152V2,       0.32) (VGG19,             0.26)
                    (Xception,          0.32)
                };
                
                \addplot[C13, fill=C13, postaction={pattern=\CThirteenPattern}] 
                coordinates {
                    (DenseNet201,       0.04) (EfficientNetV2L,   0.05)
                    (InceptionV3,       0.05) (InceptionResNetV2, 0.04)
                    (MobileNetV2,       0.05) (NASNetLarge,       0.04)
                    (RegNetY320,        0.05) (ResNet152,         0.04)
                    (ResNet152V2,       0.04) (VGG19,             0.03)
                    (Xception,          0.04)
                };
                
                \addplot[C14, fill=C14, postaction={pattern=\CFourteenPattern}] 
                coordinates {
                    (DenseNet201,       0.04) (EfficientNetV2L,   0.05)
                    (InceptionV3,       0.05) (InceptionResNetV2, 0.04)
                    (MobileNetV2,       0.05) (NASNetLarge,       0.04)
                    (RegNetY320,        0.05) (ResNet152,         0.04)
                    (ResNet152V2,       0.04) (VGG19,             0.03)
                    (Xception,          0.04)
                };

                \legend{
                    \ullpackwa{W1A1}    , \ullpackwa{W2A2}      , \ullpackwa{W3A3}  ,
                    \ourmethodwa{1}     , \ourmethodwa{2}       , \ourmethodwa{4}   ,
                    \xnnpackint{}       , \tfliteint{}          , \gemmlowpint{}    ,
                    \xnnpackfp{}        , \ruyfp{}              , \tflitefp{}       , \eigenfp{}
                }
            \end{axis}
        \end{tikzpicture}
    \end{footnotesize}
    \captionof{figure}{
        Speedup of each method against \ruyint{} on Fully Connected layers of a few well-known Convolutional Neural Networks on Raspberry Pi 4 Model B.
    }
    \label{fig:appendix:endToEnd:CNNsFCs-rapsberry-pi-4}
\end{figure*}
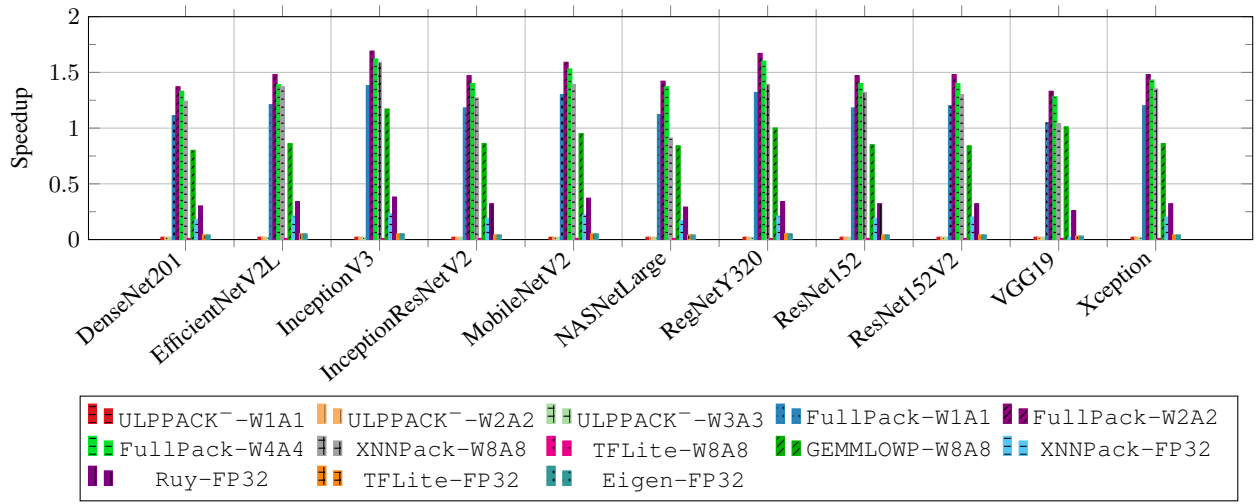